\documentclass{pic2012}

\usepackage{cite}

\def\runauthor{P. de Jong, for the ATLAS and CMS Collaborations}
\def\shorttitle{Supersymmetry searches at the LHC}

\def\ggino{\ensuremath{\mathchoice%
      {\displaystyle\raise.4ex\hbox{$\displaystyle\tilde\chi$}}%
         {\textstyle\raise.4ex\hbox{$\textstyle\tilde\chi$}}%
       {\scriptstyle\raise.3ex\hbox{$\scriptstyle\tilde\chi$}}%
 {\scriptscriptstyle\raise.3ex\hbox{$\scriptscriptstyle\tilde\chi$}}}}

\def\chinop{\ensuremath{\mathchoice%
      {\displaystyle\raise.4ex\hbox{$\displaystyle\tilde\chi^+$}}%
         {\textstyle\raise.4ex\hbox{$\textstyle\tilde\chi^+$}}%
       {\scriptstyle\raise.3ex\hbox{$\scriptstyle\tilde\chi^+$}}%
 {\scriptscriptstyle\raise.3ex\hbox{$\scriptscriptstyle\tilde\chi^+$}}}}
\def\chinom{\ensuremath{\mathchoice%
      {\displaystyle\raise.4ex\hbox{$\displaystyle\tilde\chi^-$}}%
         {\textstyle\raise.4ex\hbox{$\textstyle\tilde\chi^-$}}%
       {\scriptstyle\raise.3ex\hbox{$\scriptstyle\tilde\chi^-$}}%
 {\scriptscriptstyle\raise.3ex\hbox{$\scriptscriptstyle\tilde\chi^-$}}}}
\def\chinopm{\ensuremath{\mathchoice%
      {\displaystyle\raise.4ex\hbox{$\displaystyle\tilde\chi^\pm$}}%
         {\textstyle\raise.4ex\hbox{$\textstyle\tilde\chi^\pm$}}%
       {\scriptstyle\raise.3ex\hbox{$\scriptstyle\tilde\chi^\pm$}}%
 {\scriptscriptstyle\raise.3ex\hbox{$\scriptscriptstyle\tilde\chi^\pm$}}}}
\def\chinomp{\ensuremath{\mathchoice%
      {\displaystyle\raise.4ex\hbox{$\displaystyle\tilde\chi^\mp$}}%
         {\textstyle\raise.4ex\hbox{$\textstyle\tilde\chi^\mp$}}%
       {\scriptstyle\raise.3ex\hbox{$\scriptstyle\tilde\chi^\mp$}}%
 {\scriptscriptstyle\raise.3ex\hbox{$\scriptscriptstyle\tilde\chi^\mp$}}}}

\def\chinoonep{\ensuremath{\mathchoice%
      {\displaystyle\raise.4ex\hbox{$\displaystyle\tilde\chi^+_1$}}%
         {\textstyle\raise.4ex\hbox{$\textstyle\tilde\chi^+_1$}}%
       {\scriptstyle\raise.3ex\hbox{$\scriptstyle\tilde\chi^+_1$}}%
 {\scriptscriptstyle\raise.3ex\hbox{$\scriptscriptstyle\tilde\chi^+_1$}}}}
\def\chinoonem{\ensuremath{\mathchoice%
      {\displaystyle\raise.4ex\hbox{$\displaystyle\tilde\chi^-_1$}}%
         {\textstyle\raise.4ex\hbox{$\textstyle\tilde\chi^-_1$}}%
       {\scriptstyle\raise.3ex\hbox{$\scriptstyle\tilde\chi^-_1$}}%
 {\scriptscriptstyle\raise.3ex\hbox{$\scriptscriptstyle\tilde\chi^-_1$}}}}
\def\chinoonepm{\ensuremath{\mathchoice%
      {\displaystyle\raise.4ex\hbox{$\displaystyle\tilde\chi^\pm_1$}}%
         {\textstyle\raise.4ex\hbox{$\textstyle\tilde\chi^\pm_1$}}%
       {\scriptstyle\raise.3ex\hbox{$\scriptstyle\tilde\chi^\pm_1$}}%
 {\scriptscriptstyle\raise.3ex\hbox{$\scriptscriptstyle\tilde\chi^\pm_1$}}}}

\def\chinotwop{\ensuremath{\mathchoice%
      {\displaystyle\raise.4ex\hbox{$\displaystyle\tilde\chi^+_2$}}%
         {\textstyle\raise.4ex\hbox{$\textstyle\tilde\chi^+_2$}}%
       {\scriptstyle\raise.3ex\hbox{$\scriptstyle\tilde\chi^+_2$}}%
 {\scriptscriptstyle\raise.3ex\hbox{$\scriptscriptstyle\tilde\chi^+_2$}}}}
\def\chinotwom{\ensuremath{\mathchoice%
      {\displaystyle\raise.4ex\hbox{$\displaystyle\tilde\chi^-_2$}}%
         {\textstyle\raise.4ex\hbox{$\textstyle\tilde\chi^-_2$}}%
       {\scriptstyle\raise.3ex\hbox{$\scriptstyle\tilde\chi^-_2$}}%
 {\scriptscriptstyle\raise.3ex\hbox{$\scriptscriptstyle\tilde\chi^-_2$}}}}
\def\chinotwopm{\ensuremath{\mathchoice%
      {\displaystyle\raise.4ex\hbox{$\displaystyle\tilde\chi^\pm_2$}}%
         {\textstyle\raise.4ex\hbox{$\textstyle\tilde\chi^\pm_2$}}%
       {\scriptstyle\raise.3ex\hbox{$\scriptstyle\tilde\chi^\pm_2$}}%
 {\scriptscriptstyle\raise.3ex\hbox{$\scriptscriptstyle\tilde\chi^\pm_2$}}}}

\def\nino{\ensuremath{\mathchoice%
      {\displaystyle\raise.4ex\hbox{$\displaystyle\tilde\chi^0$}}%
         {\textstyle\raise.4ex\hbox{$\textstyle\tilde\chi^0$}}%
       {\scriptstyle\raise.3ex\hbox{$\scriptstyle\tilde\chi^0$}}%
 {\scriptscriptstyle\raise.3ex\hbox{$\scriptscriptstyle\tilde\chi^0$}}}}

\def\ninoone{\ensuremath{\mathchoice%
      {\displaystyle\raise.4ex\hbox{$\displaystyle\tilde\chi^0_1$}}%
         {\textstyle\raise.4ex\hbox{$\textstyle\tilde\chi^0_1$}}%
       {\scriptstyle\raise.3ex\hbox{$\scriptstyle\tilde\chi^0_1$}}%
 {\scriptscriptstyle\raise.3ex\hbox{$\scriptscriptstyle\tilde\chi^0_1$}}}}
\def\ninotwo{\ensuremath{\mathchoice%
      {\displaystyle\raise.4ex\hbox{$\displaystyle\tilde\chi^0_2$}}%
         {\textstyle\raise.4ex\hbox{$\textstyle\tilde\chi^0_2$}}%
       {\scriptstyle\raise.3ex\hbox{$\scriptstyle\tilde\chi^0_2$}}%
 {\scriptscriptstyle\raise.3ex\hbox{$\scriptscriptstyle\tilde\chi^0_2$}}}}
\def\ninothree{\ensuremath{\mathchoice%
      {\displaystyle\raise.4ex\hbox{$\displaystyle\tilde\chi^0_3$}}%
         {\textstyle\raise.4ex\hbox{$\textstyle\tilde\chi^0_3$}}%
       {\scriptstyle\raise.3ex\hbox{$\scriptstyle\tilde\chi^0_3$}}%
 {\scriptscriptstyle\raise.3ex\hbox{$\scriptscriptstyle\tilde\chi^0_3$}}}}
\def\ninofour{\ensuremath{\mathchoice%
      {\displaystyle\raise.4ex\hbox{$\displaystyle\tilde\chi^0_4$}}%
         {\textstyle\raise.4ex\hbox{$\textstyle\tilde\chi^0_4$}}%
       {\scriptstyle\raise.3ex\hbox{$\scriptstyle\tilde\chi^0_4$}}%
 {\scriptscriptstyle\raise.3ex\hbox{$\scriptscriptstyle\tilde\chi^0_4$}}}}

\def\squark{\ensuremath{\tilde{q}}}
\def\gluino{\ensuremath{\tilde{g}}}
\def\stop{\ensuremath{\tilde{t}}}

\def\sbottom{\ensuremath{\tilde{b}}}

\def\slepton{\ensuremath{\tilde{\ell}}}
\def\sleptonL{\ensuremath{\tilde{\ell}_{\mathrm{L}}}} % Subscript roman not italic (EE)
\def\sleptonR{\ensuremath{\tilde{\ell}_{\mathrm{R}}}} % Subscript roman not italic (EE)

\def\pt{\ensuremath{p_{\mathrm{T}}}} % Subscript roman not italic (EE)
\def\HT{\ensuremath{H_{\mathrm{T}}}} % Subscript roman not italic (EE)
\def\met{\ensuremath{E_{\mathrm{T}}^{\mathrm{miss}}}} % Sub/superscript roman not italic (EE)
\def\TeV{\ifmmode {\mathrm{\ Te\kern -0.1em V}}\else
                   \textrm{Te\kern -0.1em V}\fi}%
\def\GeV{\ifmmode {\mathrm{\ Ge\kern -0.1em V}}\else
                   \textrm{Ge\kern -0.1em V}\fi}%
\def\MeV{\ifmmode {\mathrm{\ Me\kern -0.1em V}}\else
                   \textrm{Me\kern -0.1em V}\fi}%
\def\keV{\ifmmode {\mathrm{\ ke\kern -0.1em V}}\else
                   \textrm{ke\kern -0.1em V}\fi}%
\def\eV{\ifmmode  {\mathrm{\ e\kern -0.1em V}}\else
                   \textrm{e\kern -0.1em V}\fi}%

\newcommand{\meff}{\ensuremath{m_{\mathrm{eff}}}}

\begin{document}
% \linenumbers

\title{SUPERSYMMETRY SEARCHES AT THE LHC
}

\author{P~DE~JONG}

\address{for the ATLAS and CMS Collaborations \\
Nikhef, P.O. Box 41882, 1009 DB Amsterdam, the Netherlands \\
E-mail: paul.de.jong@nikhef.nl }

\maketitle

\abstracts{Recent results in the search for supersymmetry in $pp$ collisions
at $\sqrt{s} = 7 \TeV$ and $\sqrt{s} = 8 \TeV$ by the ATLAS and CMS
experiments at the LHC are reviewed. After discussing features of inclusive
analyses and the presentation of results, emphasis will be put on searches 
for third generation squarks,
on searches for gauginos, and on possible ways supersymmetry could escape
the present analyses.
}

\section{Introduction} 

Supersymmetry (SUSY)~\cite{susyprimer} is one of the most promising and most studied theories
for physics beyond the Standard Model (SM). A supersymmetry operator changes
a fermion into a boson, and vice versa, while leaving the non spin-related
quantum numbers unchanged. Since no SUSY partners of the fundamental SM particles 
have been observed yet, SUSY, if it exists at all, must be broken.
SUSY with SUSY particles masses at the electroweak scale is particularly
interesting, as it could solve the hierarchy problem, provide a dark matter
candidate particle, and enable gauge coupling unification at the GUT scale.

It is therefore not a surprise that searches for SUSY are high priority for
the LHC experiments ATLAS~\cite{atlasdet} and CMS~\cite{cmsdet}
 (and of course indirectly also for LHCb). The
results presented in this paper are derived from analyzing typically $5$ fb$^{-1}$
of $pp$ collision data at $\sqrt{s} = 7 \TeV$ by ATLAS and CMS; some analyses
have also used up to $6$ fb$^{-1}$ of data at $\sqrt{s} = 8 \TeV$ collected in 2012.
The analysis output of ATLAS and CMS is very large, and documented in
Ref.~\cite{atlasdoc} and Ref.~\cite{cmsdoc}. In this paper only a selection
of results is presented, highlighting recent developments. After a
discussion of inclusive searches, emphasis will shift to searches for stops
and sbottoms, as well as searches for gauginos. Possible ways SUSY might have
escaped detection by these analyses are discussed, and a number of ``non-standard''
analyses are described. Finally, an outlook will be given.

\section{Inclusive Searches}

SUSY production at the LHC will be dominated by squark-pair, gluino-pair and
squark-gluino production, if squarks and gluinos are not excessively massive.
Squarks and gluinos are expected to decay, typically promptly, in short or
long decay chains that end up with the production of the lightest SUSY
particle (LSP), which is expected to be stable if R-parity is 
conserved~\cite{susyprimer}. In many
models this particle is the lightest neutralino or the gravitino, 
and by the fact that it only interacts very weakly, events are characterized 
by large missing transverse momentum ($\met$). 
Further typical characteristics are energetic jets, and possibly isolated
leptons from gaugino or $W/Z$ boson decay. Backgrounds are dominated by multi-jet
production, either with real missing momentum or with mis-measured jets,
top quark pair production, and $W/Z$ production with associated jets.

\begin{figure}[!thb]
\begin{center}
\includegraphics[width=6cm]{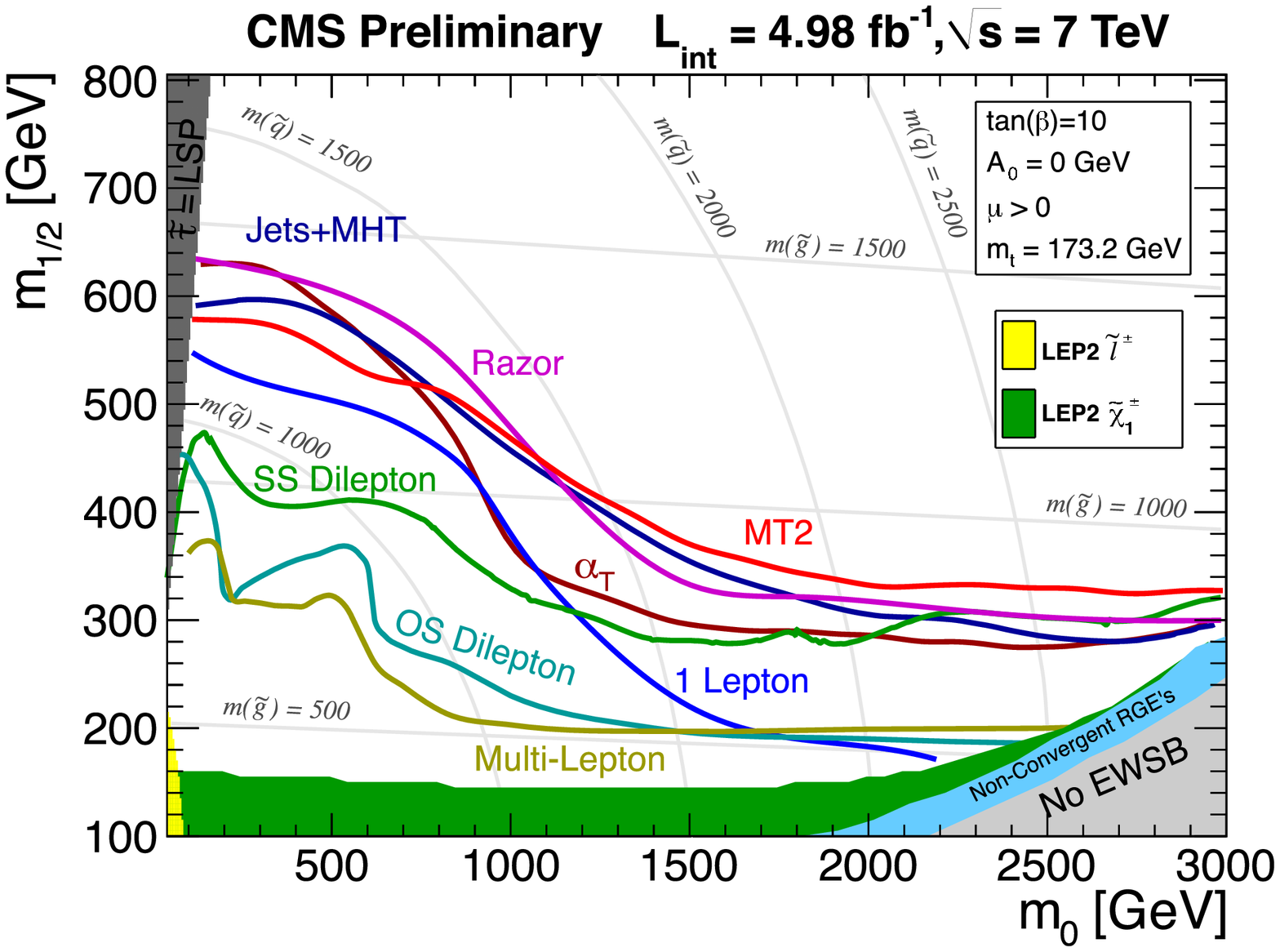}
\includegraphics[width=6cm]{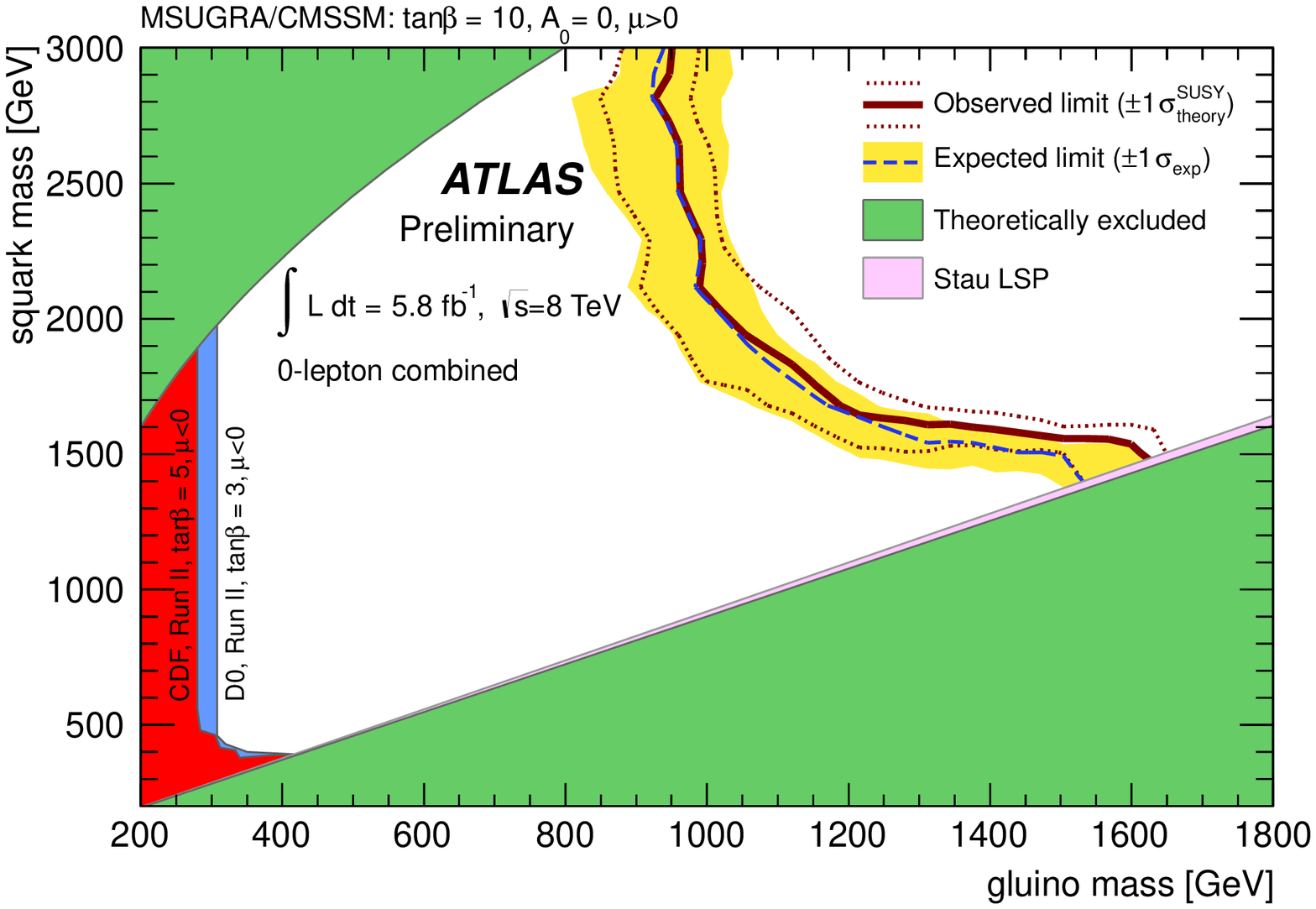}
\caption[*]{Left: A summary of CMS exclusion limits in the MSUGRA/CMSSM 
$m_0,m_{1/2}$ plane~\cite{cmsdoc}. 
Right: MSUGRA/CMSSM exclusion limits of ATLAS expressed in terms of squark
and gluino masses~\cite{atlasjetsmet}.
}
\label{fig:msugra}
\end{center}
\end{figure}

Both ATLAS and CMS have developed analyses targeted towards effective
suppression of backgrounds as well as reliable estimation of whatever
background remains, while keeping the analyses general and not overtuned
to any specific model.
CMS uses a number of variables that provide separation
between signal and background, such as $\alpha_{\mathrm{T}}$~\cite{alphat,cmsalphat}, 
the razor variables~\cite{cmsrazor}, the scalar sum of $\pt$ of all jets $\HT$
and missing $\HT$~\cite{cmsjetsmht}, and the stransverse mass 
$M_{\mathrm{T2}}$~\cite{mt2,cmsmt2}. ATLAS mostly uses the effective mass
variable $\meff$, defined as the $\pt$ sum of jets, $\met$ and possibly
leptons~\cite{atlasjetsmet}, but also other variables, such as
a variable related to the missing transverse momentum significance, 
$\met/\sqrt{\HT}$~\cite{atlasmultijet}. 
Many results on multiple final states with no, one, two or more leptons
have been published~\cite{cmsdoc,atlasdoc}.
None of the searches have observed a significant excess above the SM
prediction, and limits on new physics have been set. As an example, 
Fig.~\ref{fig:msugra} shows the derived exclusion limits in the MSUGRA/CMSSM
framework in the $m_0,m_{1/2}$ plane (CMS, left) or in the squark-gluino
mass plane (ATLAS, right) for $\tan \beta = 10$, $A_0 = 0$,
and $\mu > 0$. Clearly these results provide significant constraints on
SUSY particle masses in the MSUGRA/CMSSM model: for all squark masses,
gluinos with a mass below $950 \GeV$ are excluded, and for all gluino
masses, squarks with a mass below $1400 \GeV$ are excluded.

\section{Simplified Models}

The general MSSM adds $105$ new parameters to the parameter set of the Standard
Model~\cite{susyprimer}; these $105$ parameters reflect our ignorance on how SUSY is broken.
Such a large parameter set is hard to work with in experimental and theoretical
analyses. Making ever stronger assumptions can reduce the parameter set~\cite{susypdg}: 
the {\it phenomenological} MSSM (CP-conserving MSSM with minimal flavour violation
and degenerate first two sfermion generations) has $19$ parameters, and
top-down models like MSUGRA/CMSSM or minimal gauge mediation (GMSB) have only
a handful, which make them easy to work with. However, the assumptions made in
these models could well be unfounded, leading to incorrect conclusions on SUSY.

In the past years, the use of simplified models in SUSY searches has gained
ground~\cite{simplifiedmodels}. 
A simplified model is defined by an effective Lagrangian describing
the interactions of a small number of new particles. Simplified models can
equally well be described by a small number of masses and cross sections.
A typical simplified model assumes one production channel, and one decay channel
with a 100\% branching fraction (although this is not a strict requirement), in
circumstances where other particles are very heavy and decouple. Simplified
models are useful to study where an analysis is effective and where it is
not, but also to present results, and characterize signals of new physics
if such signals were found.

\begin{figure}[!thb]
\begin{center}
\includegraphics[width=6cm]{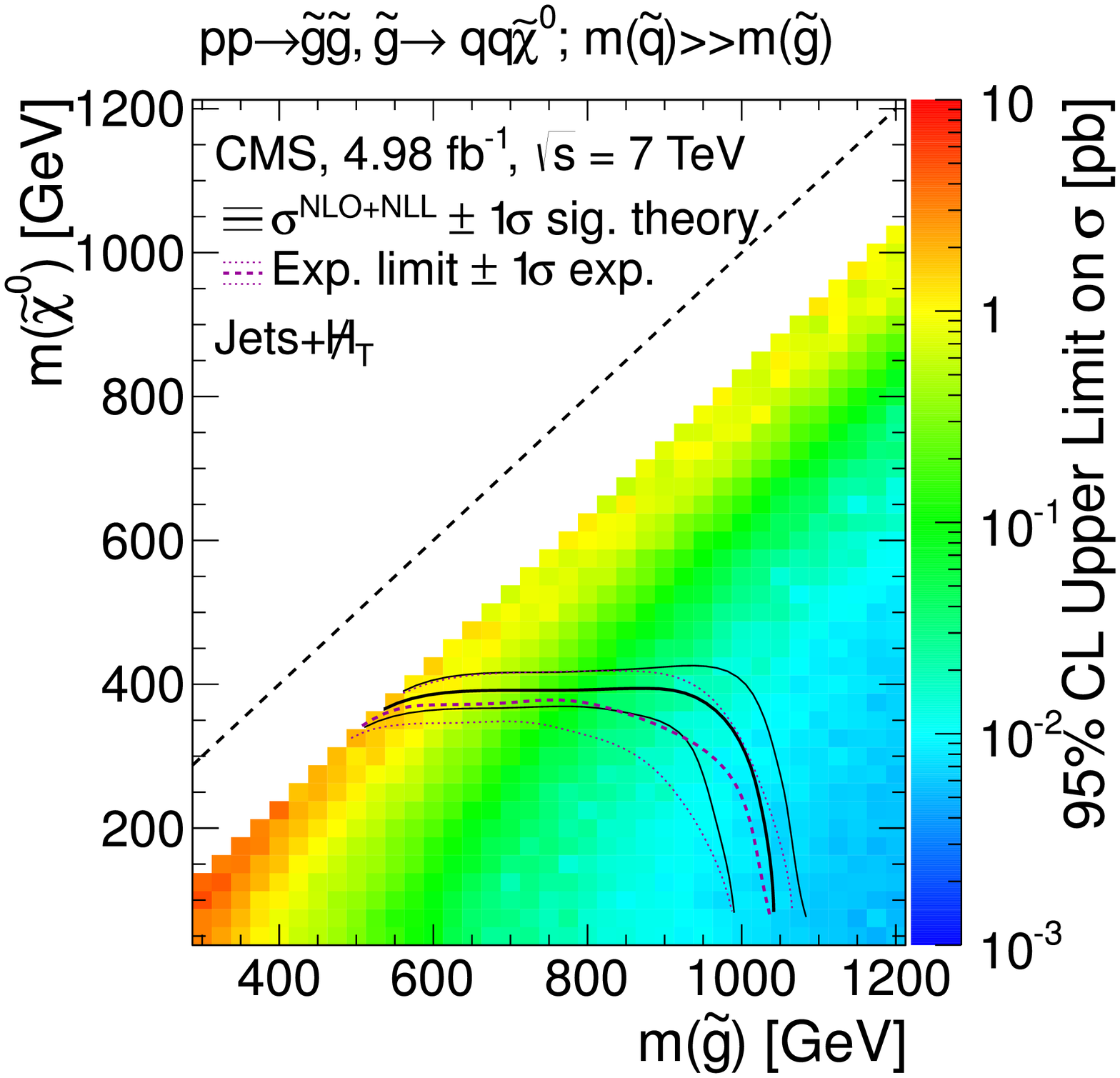}
\includegraphics[width=6cm]{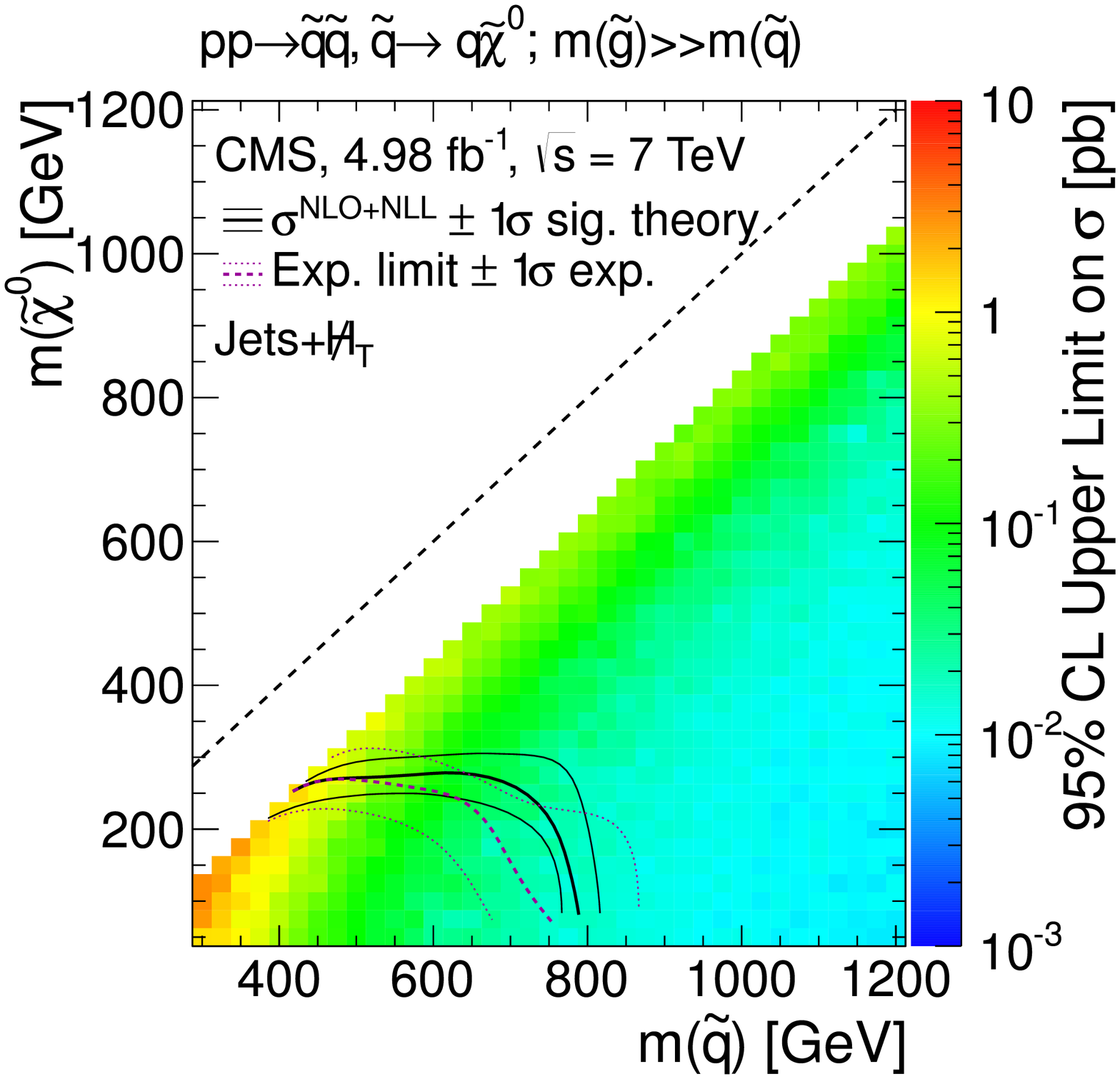}
\includegraphics[width=6cm]{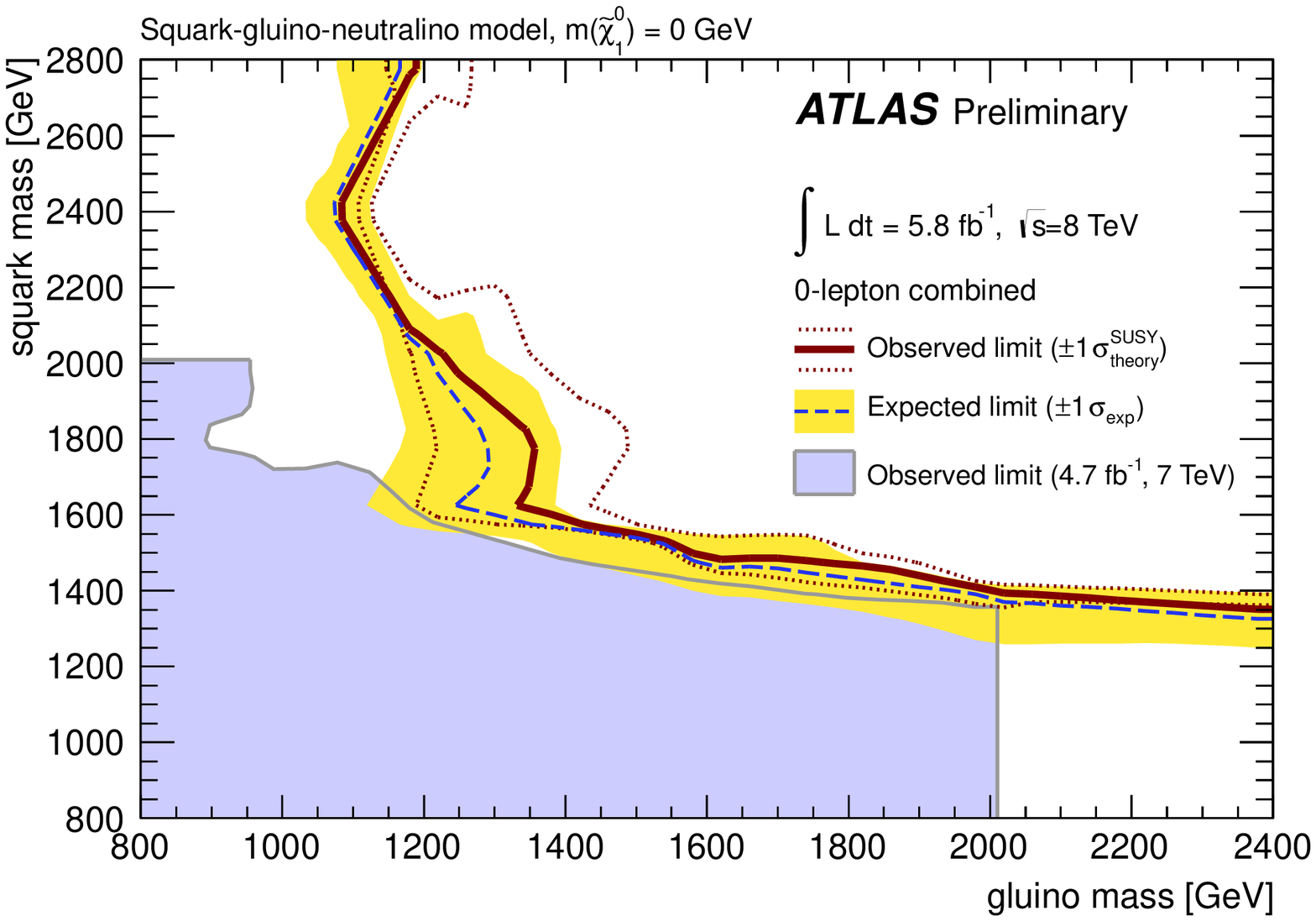}
\includegraphics[width=6cm]{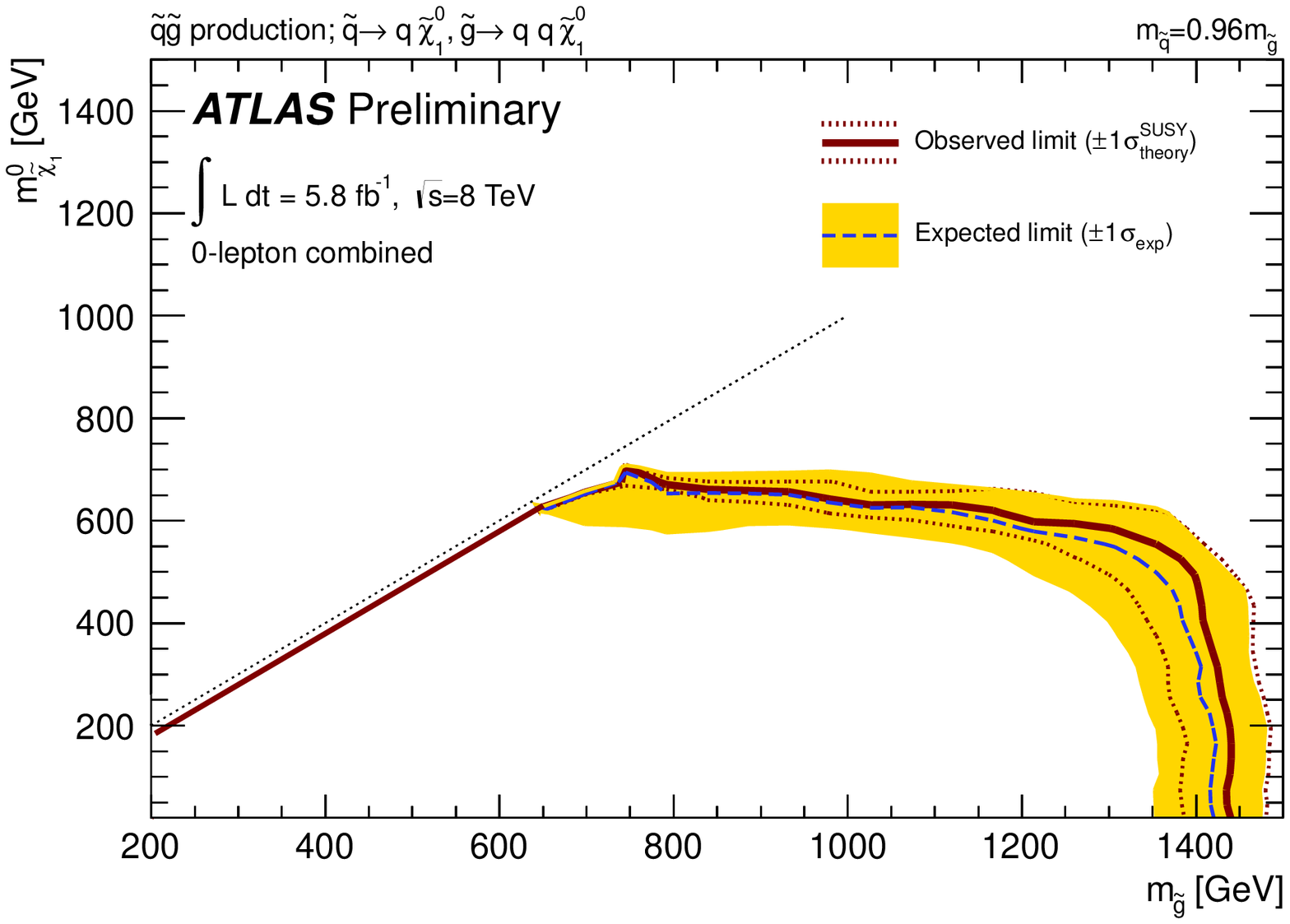}
\caption[*]{Top: CMS exclusion limits in simplified models with
only $\gluino \to q \bar{q} \ninoone$ (left) or only
$\squark \to q \ninoone$ (right)~\cite{cmsjetsmht}.
Bottom: ATLAS exclusion limits in simplified models with only
strong production of gluinos and first- and second generation
squarks, with direct decays to jets and massless neutralinos
(left), or only squark+gluino production, with 
$\squark \to q \ninoone$, and $\gluino \to q \bar{q} \ninoone$,
and assuming $m_{\squark} = 0.96 m_{\gluino}$ (right)~\cite{atlasjetsmet}.}
\label{fig:simplified}
\end{center}
\end{figure}

As an example, Fig.~\ref{fig:simplified} shows simplified model results of
inclusive searches for squarks and gluinos in the jets plus $\met$
final state by CMS~\cite{cmsjetsmht} and ATLAS~\cite{atlasjetsmet}. 
The models used assume specific decay modes with 100\% branching
fraction: $\gluino \to q \bar{q} \ninoone$ and $\squark \to q \ninoone$,
and specific production modes: gluino pair production, squark pair production,
or squark plus gluino production.
These plots show that limits on squark and gluino masses in the $\TeV$ scale are
in fact only valid for not too large neutralino masses, below $300-400 \GeV$,
and that these limits collapse for heavier neutralinos.
Many more simplified model interpretations are available, also with
more complicated decay modes~\cite{cmsdoc,atlasdoc}.
CMS has published a dedicated note with simplified model interpretations
of a number of analyses~\cite{cmssimplified}.

\section{Gauge mediation}

In gauge-mediated models of supersymmetry breaking, the breaking is mediated
from the invisible sector where breaking takes place to the electroweak scale
by gauge interactions. The phenomenology of gauge mediated SUSY breaking models is
characteristic: a very light gravitino is LSP, and a number of candidates can
be the next-to-lightest SUSY particle (NLSP), such as the stau, other sleptons, 
or the neutralino. NLSP decay will
then lead, respectively, to tau plus missing momentum, lepton plus missing
momentum, or photon/$Z$/Higgs plus missing momentum final states; the latter
fractions being determined by the bino/wino/higgsino nature of the neutralino.

\begin{figure}[!thb]
\begin{center}
\includegraphics[width=6cm]{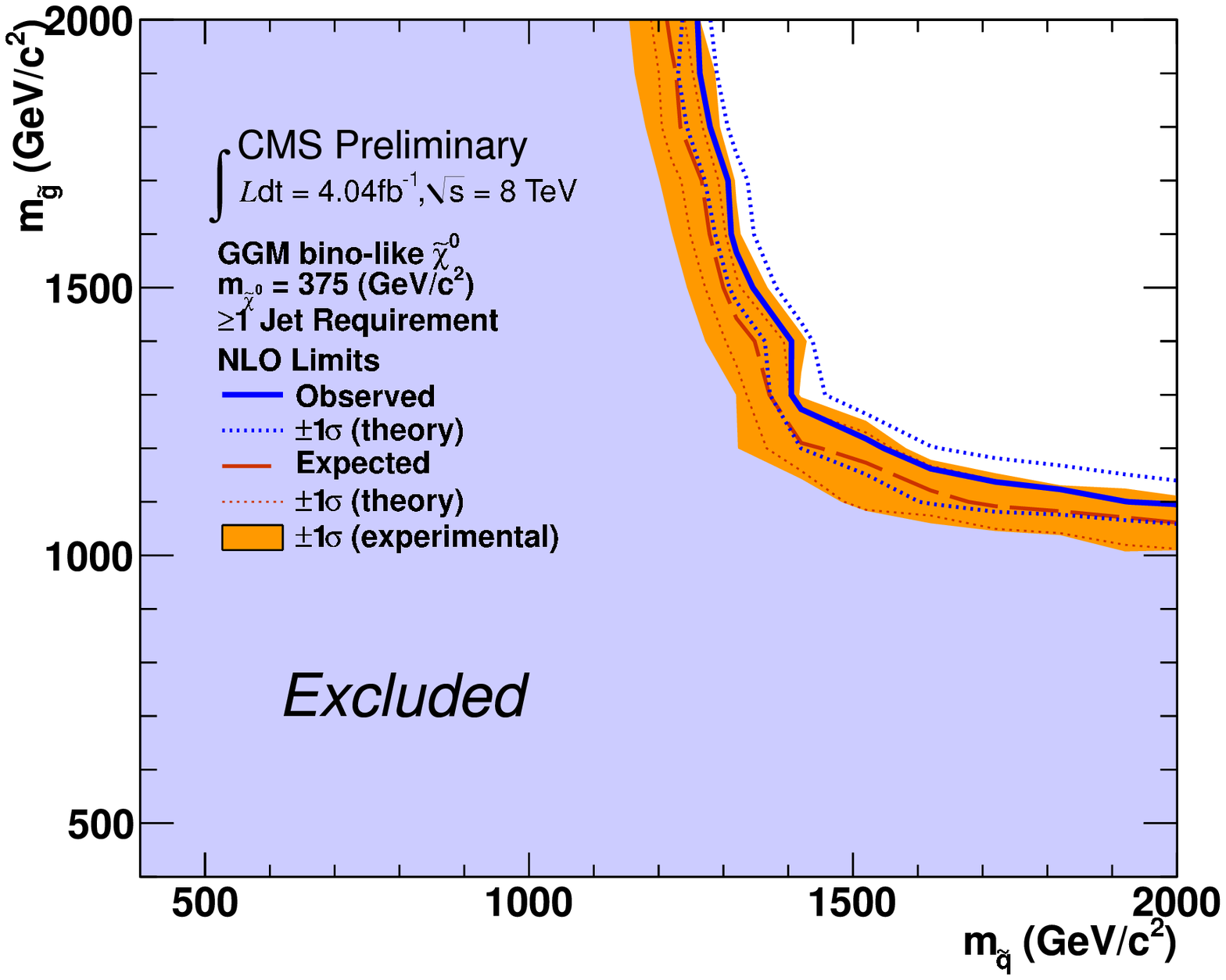}
\includegraphics[width=6cm]{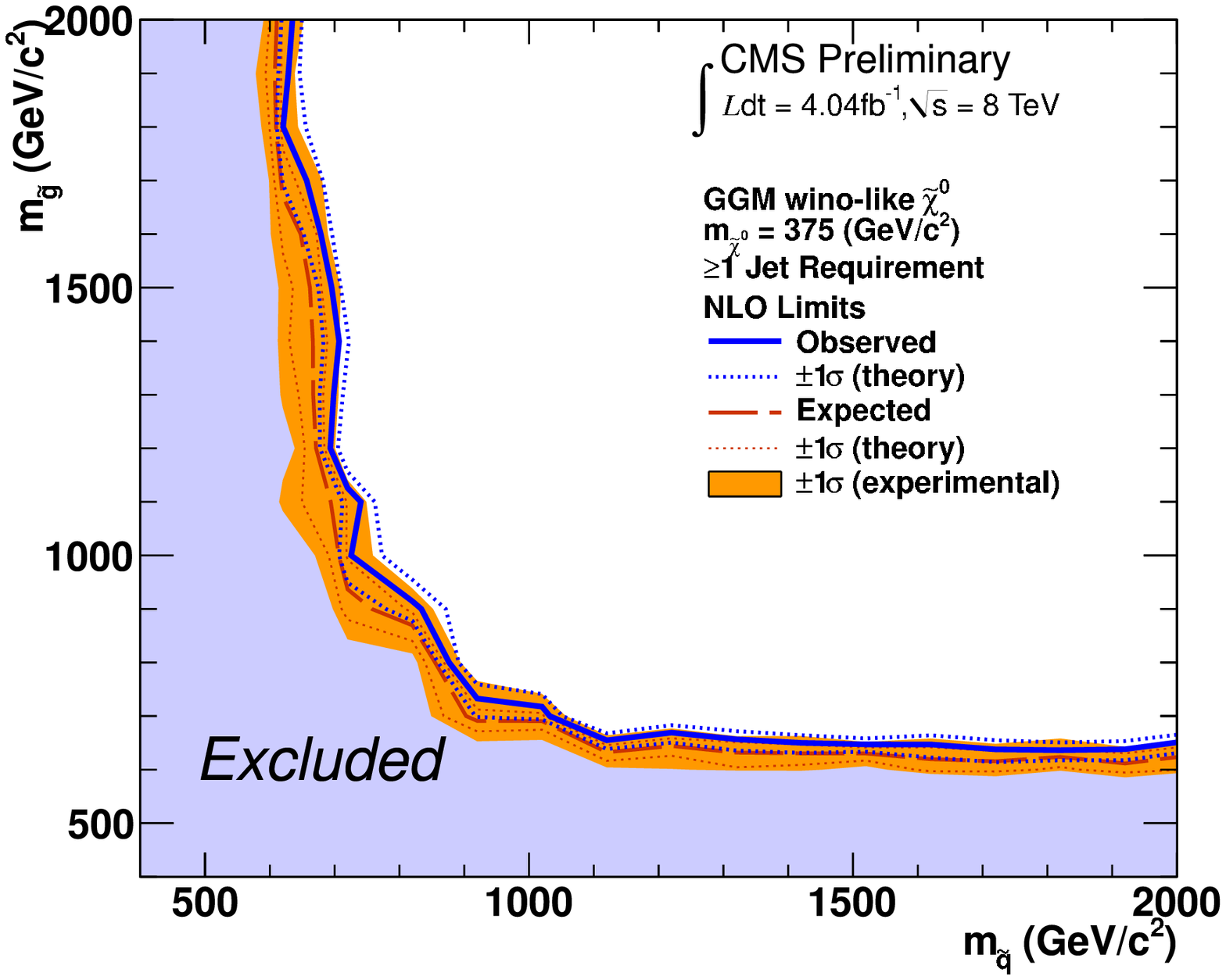}
\includegraphics[width=6cm]{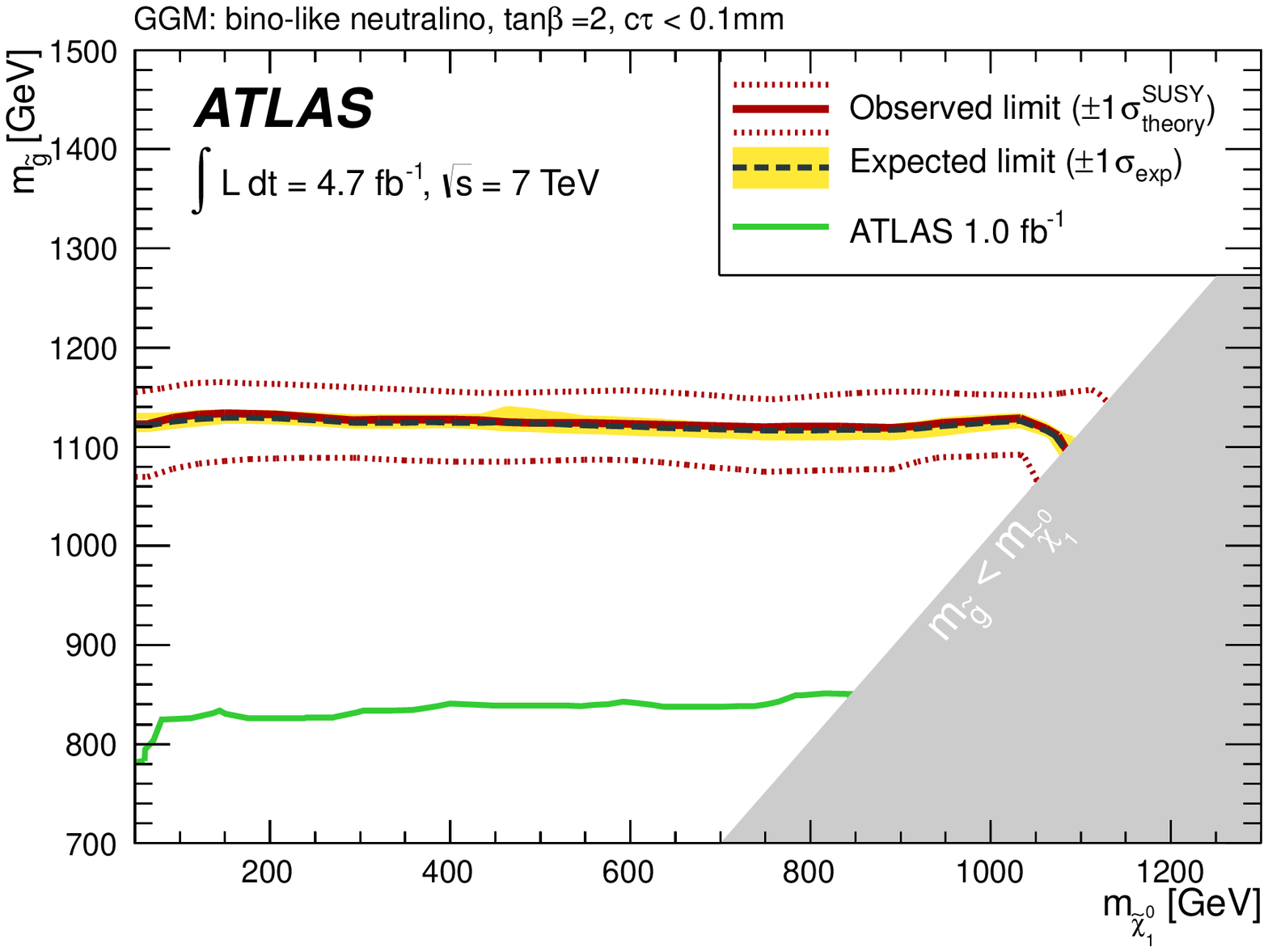}
\includegraphics[width=6cm]{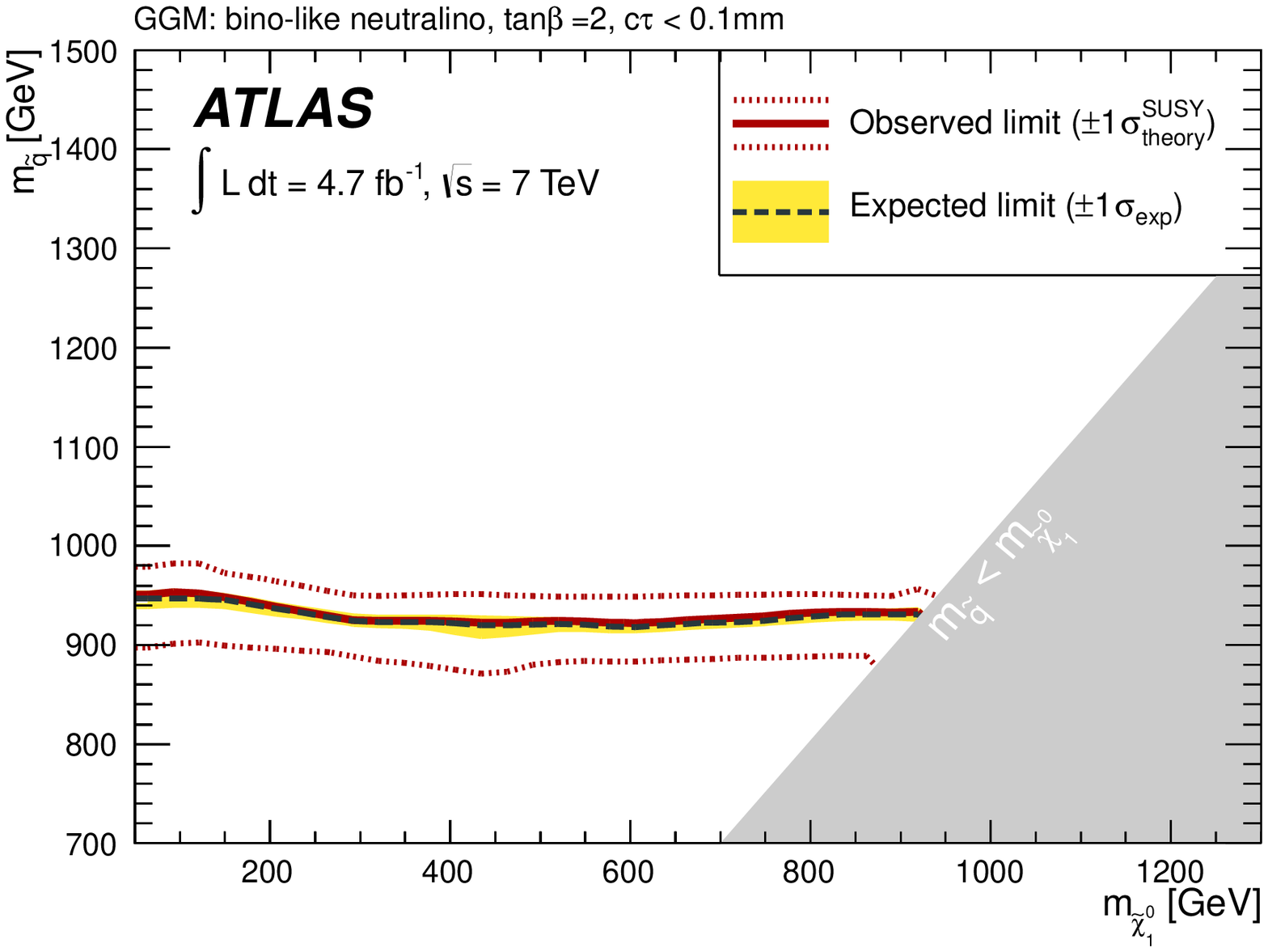}
\caption[*]{Top: CMS limits in general gauge mediation for bino-like
neutralinos (left) and wino-like neutralinos (right) from the
diphoton plus $\met$ analysis~\cite{cmsdiphoton}. 
Bottom: ATLAS limits in GGM
for bino-like neutralinos, for gluino production (left) and squark
production (right)~\cite{atlasdiphoton}.}
\label{fig:ggm}
\end{center}
\end{figure}

Both ATLAS and CMS have searched for such final states in a number of
analyses. As an example, Fig.~\ref{fig:ggm} shows limits in general
gauge mediation (GGM) models resulting from analyses of diphoton plus missing
transverse momentum final states~\cite{cmsdiphoton,atlasdiphoton}.
For bino-like neutralinos, squark and gluino masses below the $\TeV$ scale
are excluded in GGM, regardless of the neutralino mass. Limits for wino-like
neutralinos are somewhat lower.

\section{Naturalness}

The results of inclusive searches for SUSY at the LHC put ever
tighter constraints. In MSUGRA/CMSSM, the limits on squark and gluino
masses well exceed $1 \TeV$, and 
also in simplified models, the mass limits on squarks of the first and
second generation and on gluinos are of order $1 \TeV$, at least for
not too massive neutralinos.

However, reflecting on the actual merits of supersymmetry at
the electroweak scale, one could conclude that some SUSY particles may
actually be more important than others. In particular,
guided by the hierarchy problem and by the dark matter problem,
such reflections may proceed as follows~\cite{susyprimer}:
\begin{itemize}
\item Hierarchy problem and the Higgs boson mass: the dominant 
radiative corrections to the Higgs boson mass in the SM
arise from the top quark loop. In SUSY, these corrections are
balanced by a top squark loop. A relatively light stop will ensure
good cancellations, and the need of only small remaining ``fine-tuning''.
\item Fine-tuning and the $\mu$ term: in SUSY, the mass of the $Z$ boson 
is related to the Higgsino mass term $\mu$. A ``natural'' scale for
$\mu$, without the need for excessive fine-tuning, is thus the
electroweak scale. The masses of the gauginos are partly determined 
by the value of $\mu$, and are therefore naturally of the electroweak scale.
\item Dark matter: the existence of a large non-baryonic component
to mass in the universe is now well established, and a lightest
stable SUSY particle in the $1 \GeV-1 \TeV$ mass range would be
an excellent dark matter candidate.
\end{itemize}

One could thus arrive at a picture of ``natural'' SUSY, with
low or acceptable levels of fine-tuning, with stops below
$\sim 1 \TeV$, gluinos (which affect the stop mass) below $\sim 2 \TeV$,
and gauginos with masses of order $\sim 200 \GeV$ or so.
In the following sections, dedicated searches for third generation
squarks and for gauginos will be discussed.

\section{Third Generation Squarks Searches}

Stops ($\stop$) and sbottoms ($\sbottom$) can be produced in 
the decay of gluinos,
or they can be produced directly in pairs. Production via gluino
decay will dominate if gluinos are not too heavy and if the
gluino decay branching fractions into stops or sbottoms are
high. Direct production of pairs of top squarks
is suppressed with respect to first and second generation squarks
by more than an order of magnitude, and there is a large
background, in particular from top-quark pair production.

Gluino-mediated on-shell sbottom production, $\gluino \to
\sbottom \bar{b}$, is possible if the gluino is heavier than the 
sbottom plus the $b$-quark\footnote{In all decays, charge-conjugate
decay modes are implied}. The sbottom could decay to 
$b + \nino$, or to $t + \chinom$. If the sbottom is too
heavy for $\gluino \to \sbottom \bar{b}$, mediation of the
gluino decay by virtual sbottoms could still lead to the final
state $b \bar{b} \nino$ (c.q. $t \bar{b} \chinom$). 
Gluino-mediated stop production can be described 
in an analogous way: $\gluino \to \stop \bar{t}$, if kinematically
allowed, with $\stop \to t \nino$ or $\stop \to b \chinop$,
or $\gluino \to t \bar{t} \nino$ (c.q. $\gluino \to t \bar{b}
\chinom$) otherwise.

\begin{figure}[!thb]
\begin{center}
\includegraphics[width=6cm]{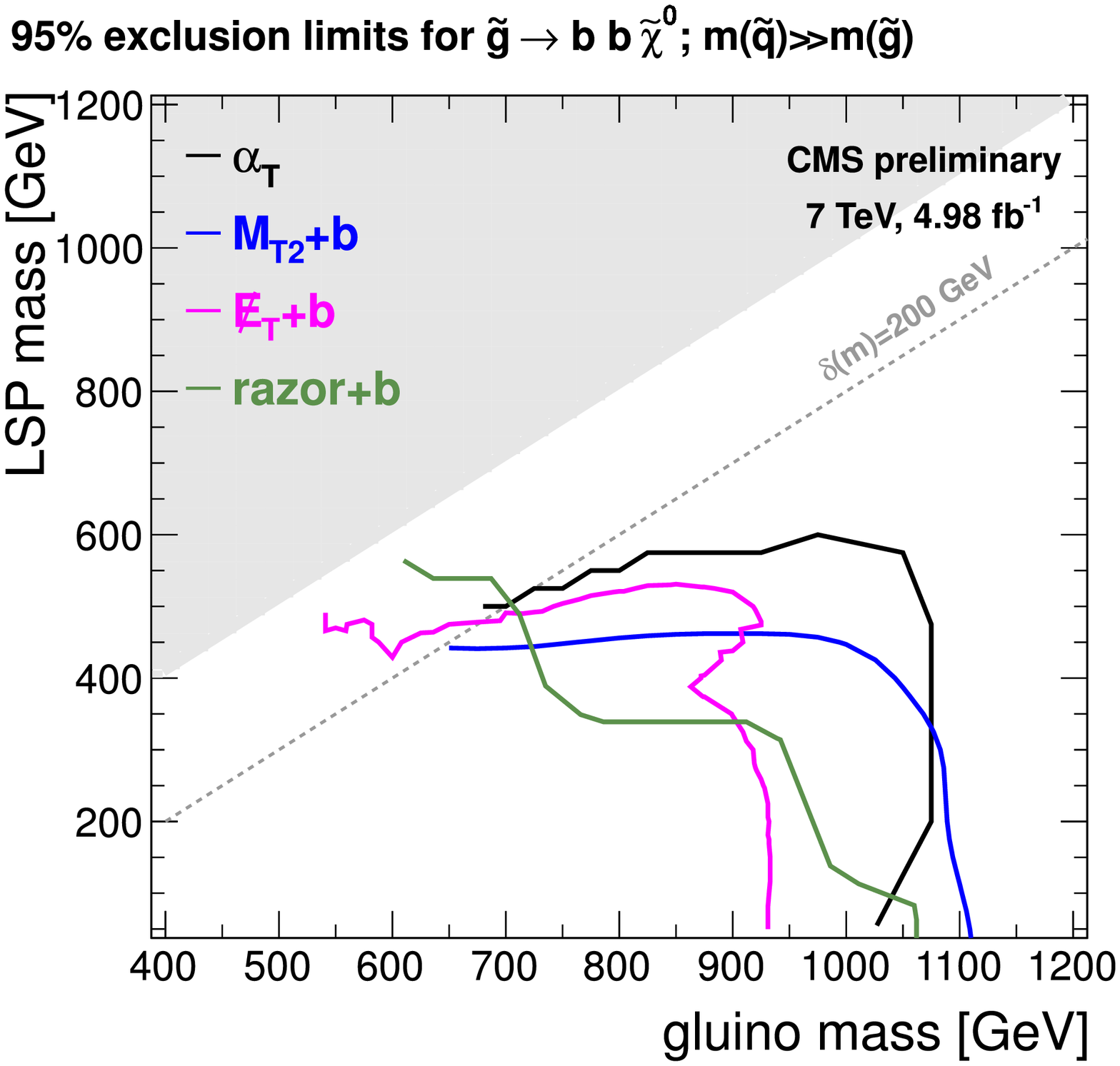}
\includegraphics[width=6cm]{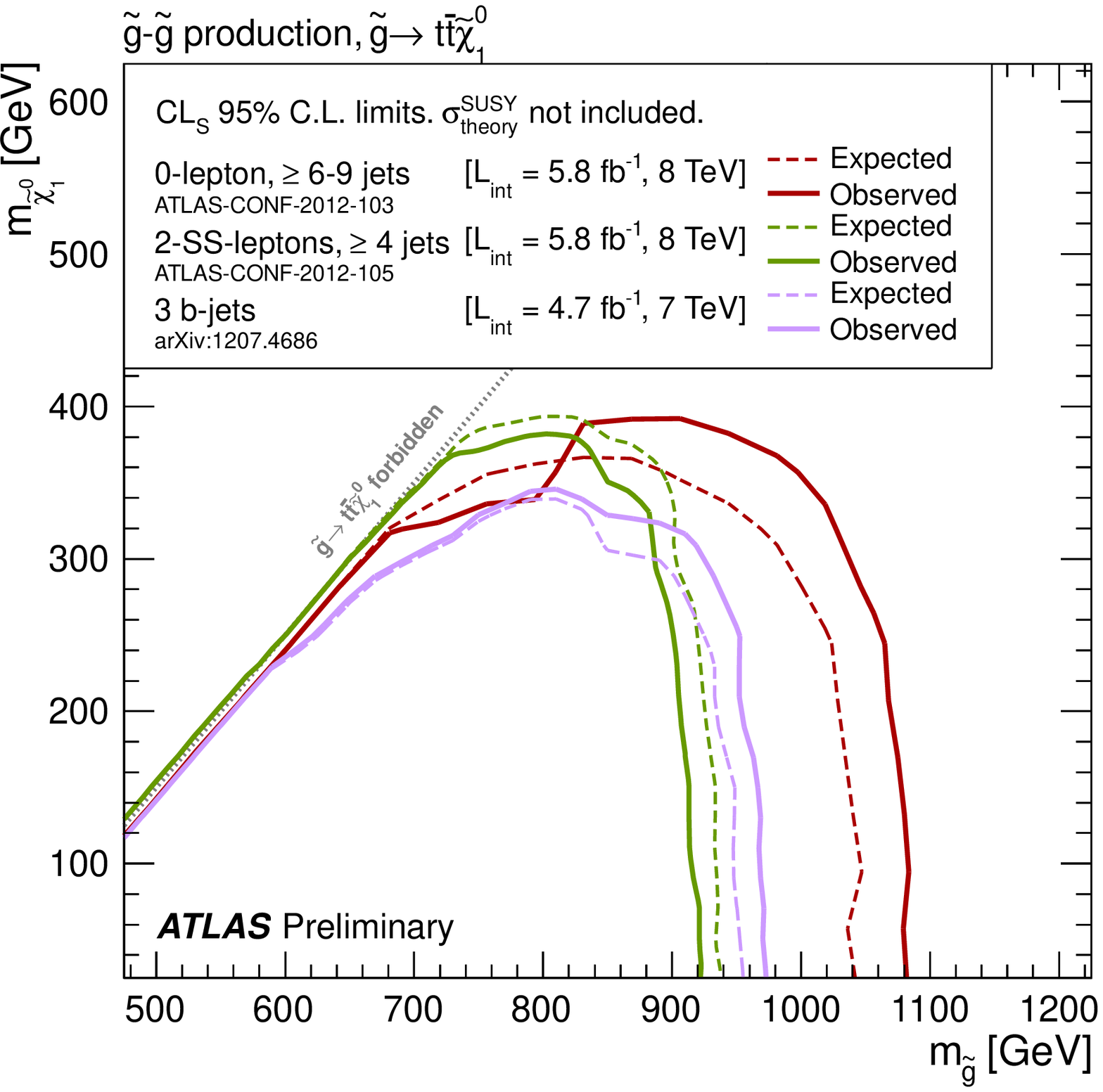}
\caption[*]{Limits in simplified models on $\gluino \to b \bar{b} \nino$
by CMS~\cite{cmsalphat,cmsmt2,cmsmetb,cmsrazorb} 
(left) and on $\gluino \to t \bar{t} \nino$
by ATLAS~\cite{atlasmultijet,atlasssjets,atlas3bjets} 
(right).}
\label{fig:gtt}
\end{center}
\end{figure}

A selection of
results from CMS~\cite{cmsalphat,cmsmt2,cmsmetb,cmsrazorb}
and ATLAS~\cite{atlasmultijet,atlasssjets,atlas3bjets}
is given in Fig.~\ref{fig:gtt}. In Fig.~\ref{fig:gtt} (left),
CMS results on $\gluino \to b \bar{b} \nino$ are shown as
limits in the gluino-neutralino mass plane; gluinos up to
almost $1.1 \TeV$ are excluded for neutralino masses up
to $500 \GeV$, assuming a 100\% branching fraction in
this mode. In Fig.~\ref{fig:gtt} (right) analogous results
for $\gluino \to t \bar{t} \nino$ obtained by ATLAS are shown,
with gluino mass limits up to $1 \TeV$ for neutralino masses
up to $350 \GeV$, again assuming 100\% branching fraction.
% CMS has results similar to ATLAS on $\gluino \to t \bar{t} \nino$,
% and ATLAS has similar results to CMS on $\gluino \to b \bar{b} \nino$.

\begin{figure}[!thb]
\begin{center}
\includegraphics[width=6cm]{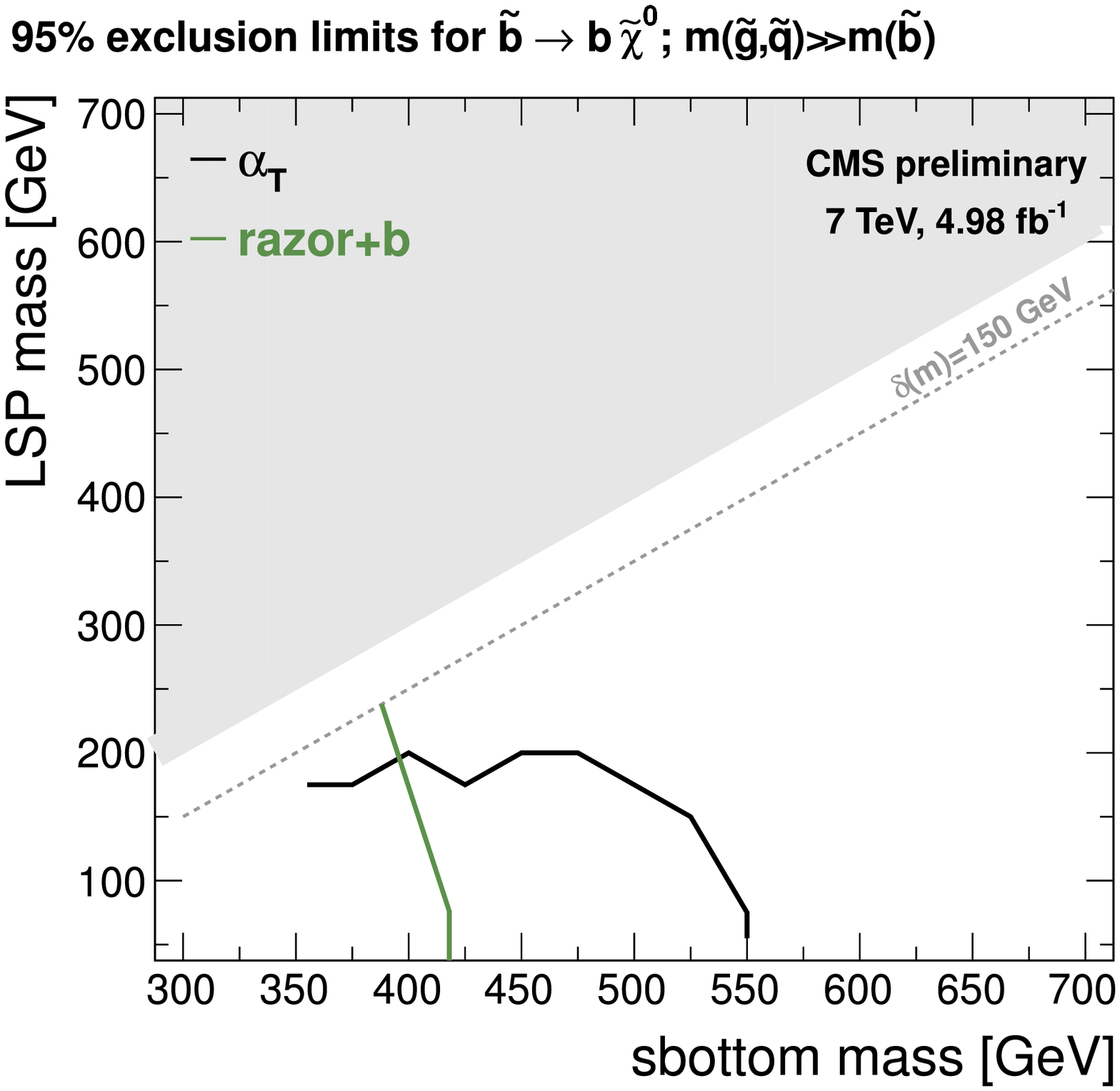}
\includegraphics[width=6cm,height=5cm]{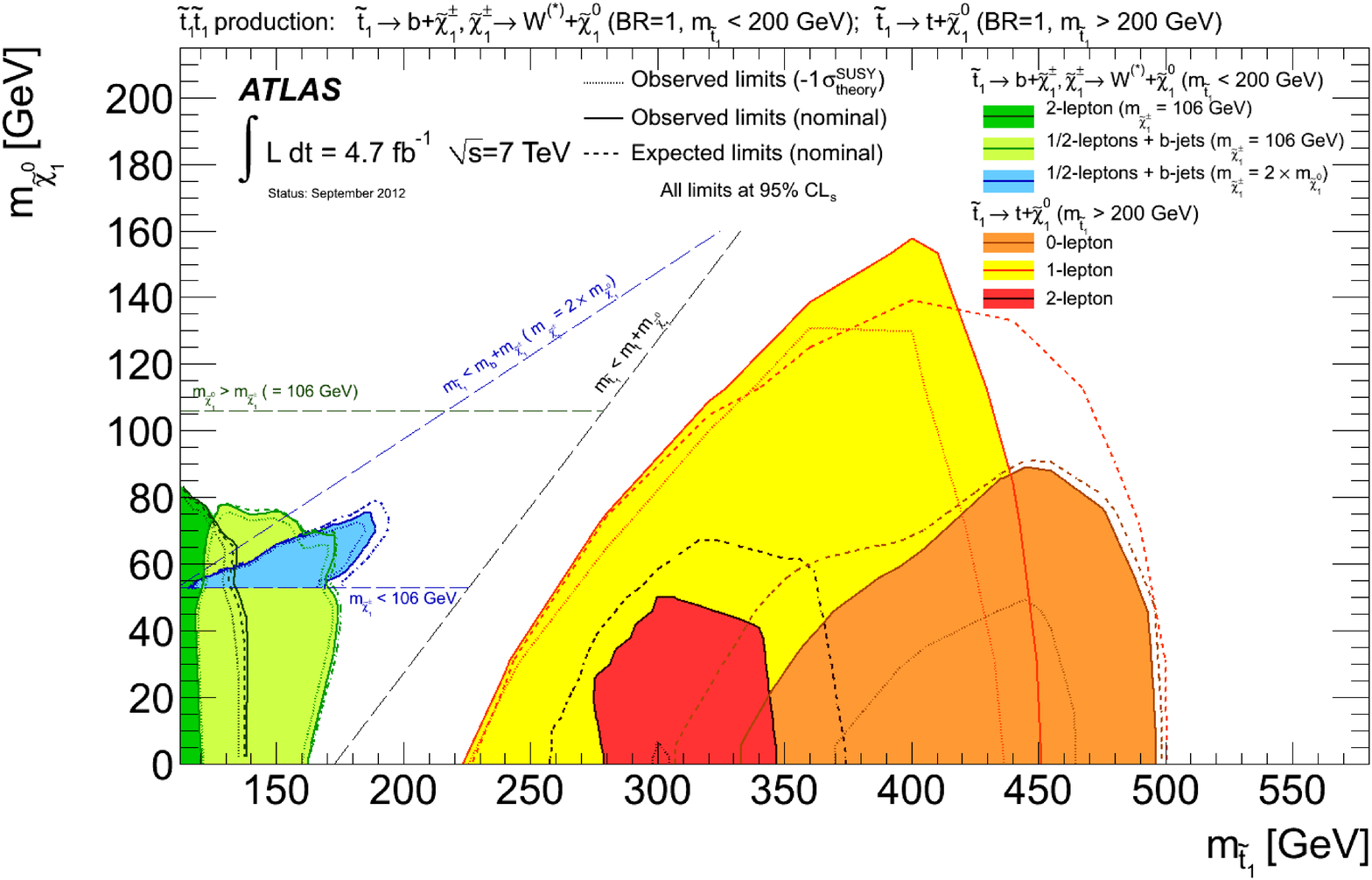}
\caption[*]{CMS limits on direct sbottom pair
production~\cite{cmsalphat,cmsrazorb} (left),
and ATLAS limits on direct stop pair
production~\cite{atlasstop1,atlasstop2,atlasstop3,atlasstop4,atlasstop5}
(right).
}
\label{fig:allstop}
\end{center}
\end{figure}

Searches for direct sbottom pair production and direct stop 
pair production are also carried out by ATLAS and CMS.
CMS have interpreted a number of inclusive analyses with
$b$-tags in terms of limits on direct sbottom pair
production~\cite{cmsalphat,cmsrazorb} as shown in
Fig.~\ref{fig:allstop} (left) and in terms of limits on
direct stop pair production~\cite{cmsalphat,cmsrazor,cmsrazorb}.
ATLAS has a dedicated analysis searching for sbottom
pair production~\cite{atlassbottom}, and five dedicated searches
for direct stop pair production~\cite{atlasstop1,atlasstop2,atlasstop3,atlasstop4,atlasstop5}, 
targeting light or heavy stops in decay
modes with no, one or two leptons. The limits derived from these
five searches are summarized in Fig.~\ref{fig:allstop} (right).
Direct sbottom pair production is excluded for sbottom masses
below $500 \GeV$ for LSP masses below $150 \GeV$, assuming
100\% sbottom decay into bottom plus neutralino. Exclusion limits on
the stop mass from
direct stop pair production reach $500 \GeV$ for massless LSPs,
and assuming specific decay modes ($\stop \to b \chinopm$ for
light stop, $\stop \to t \nino$ for heavy stop),
but the stop is not excluded in a window around the top quark
mass. 
% Again, ATLAS and CMS results on both channels are similar.

In conclusion, it is clear that there is sensitivity for third
generation squark searches, and that first relevant limits
have been set. More results can be expected, and with more
data in store this will be an exciting search channel.

\section{Gauginos}

The SUSY gauginos, i.e. the charginos $\chinopm$ and 
neutralinos $\nino$, are
mixtures of the partners of the SM gauge bosons and the
partners of the Higgs bosons, the Higgsinos. Their masses
and their bino-wino-higgsino composition are determined by
a limited number of parameters: $\mu$, $M_1$, $M_2$ and
$\tan \beta$~\cite{susyprimer}. 

The lightest neutralino $\ninoone$ is the LSP in a large
part of parameter space of many SUSY models, and is a 
good dark matter candidate. Pair production of the
lightest neutralinos at the LHC leads to an invisible final
state, but could be tagged if an extra jet or photon is
radiated. Such monojet or monophoton analyses can be interpreted
in terms of dark matter production models, as discussed
elsewhere in these proceedings~\cite{exotictalk}.

\begin{figure}[!thb]
\begin{center}
\includegraphics[angle=270,width=5cm]{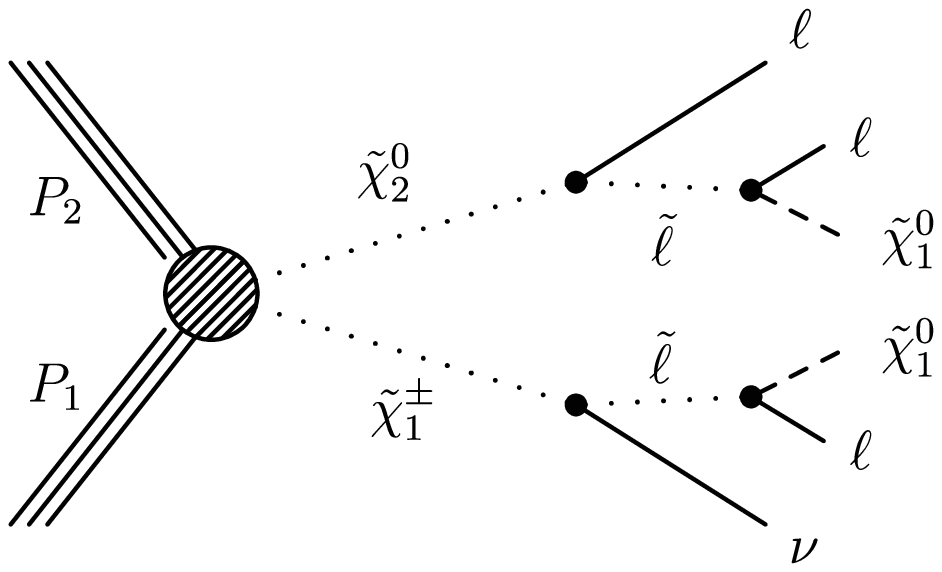}
\includegraphics[angle=270,width=5cm]{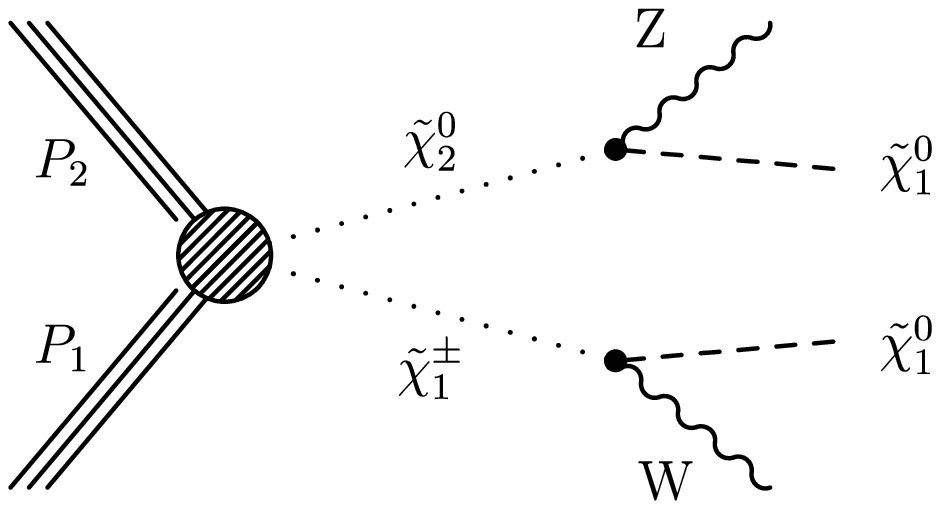}
\caption[*]{Simplified models for gaugino production and decay.
Left: decays mediated by sleptons boost the leptonic branching
fractions of gauginos. Right: if sleptons are too massive, decays
take place via real or virtual $W$ or $Z$ boson emission.
}
\label{fig:sm_gauginos}
\end{center}
\end{figure}

Heavier neutralinos, and charginos, decay into SM particles
and eventually the LSP. Since production cross sections for
gauginos are small and backgrounds for hadronic decays are
large, analyses focus on the leptonic decay modes of gauginos. 
The branching fractions of gauginos into leptons is model-dependent, 
and depends in particular on whether light sleptons can mediate 
the decay, in
which case leptonic branching fractions can be very high, or
whether sleptons are too massive, and the decay takes place
through emission of real or virtual $W$ or $Z$ bosons, 
with smaller leptonic
branching fractions. These scenarios are illustrated in
Fig.~\ref{fig:sm_gauginos}. The light slepton scenario can further
be split in a light $\sleptonL$ scenario, with equal branching
fractions in charged leptons of all flavours and neutrinos, and
a light $\sleptonR$ scenario, with dominant decay to $\tau$ leptons.

\begin{figure}[!thb]
\begin{center}
\includegraphics[width=6cm]{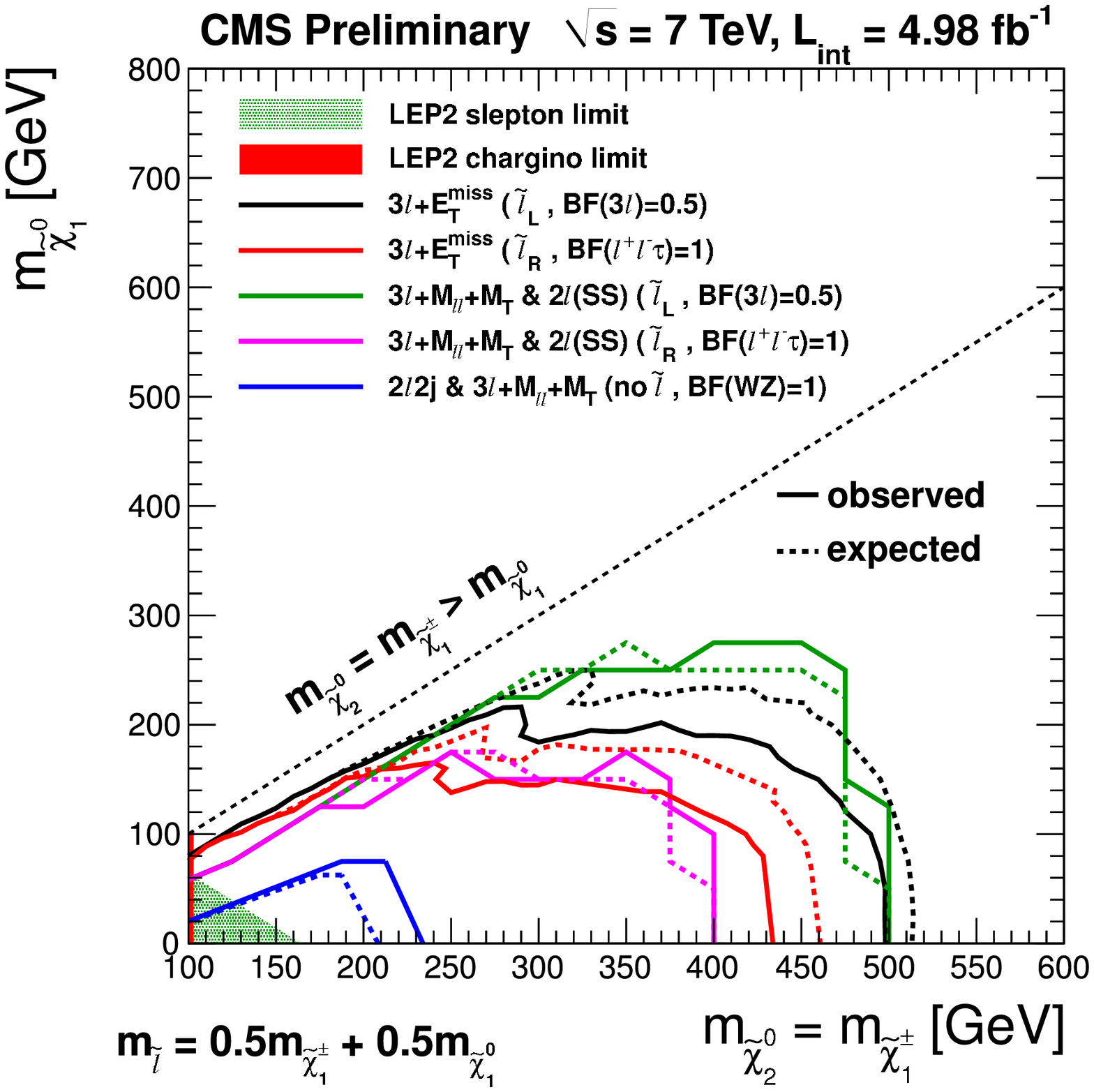}
\includegraphics[width=6cm]{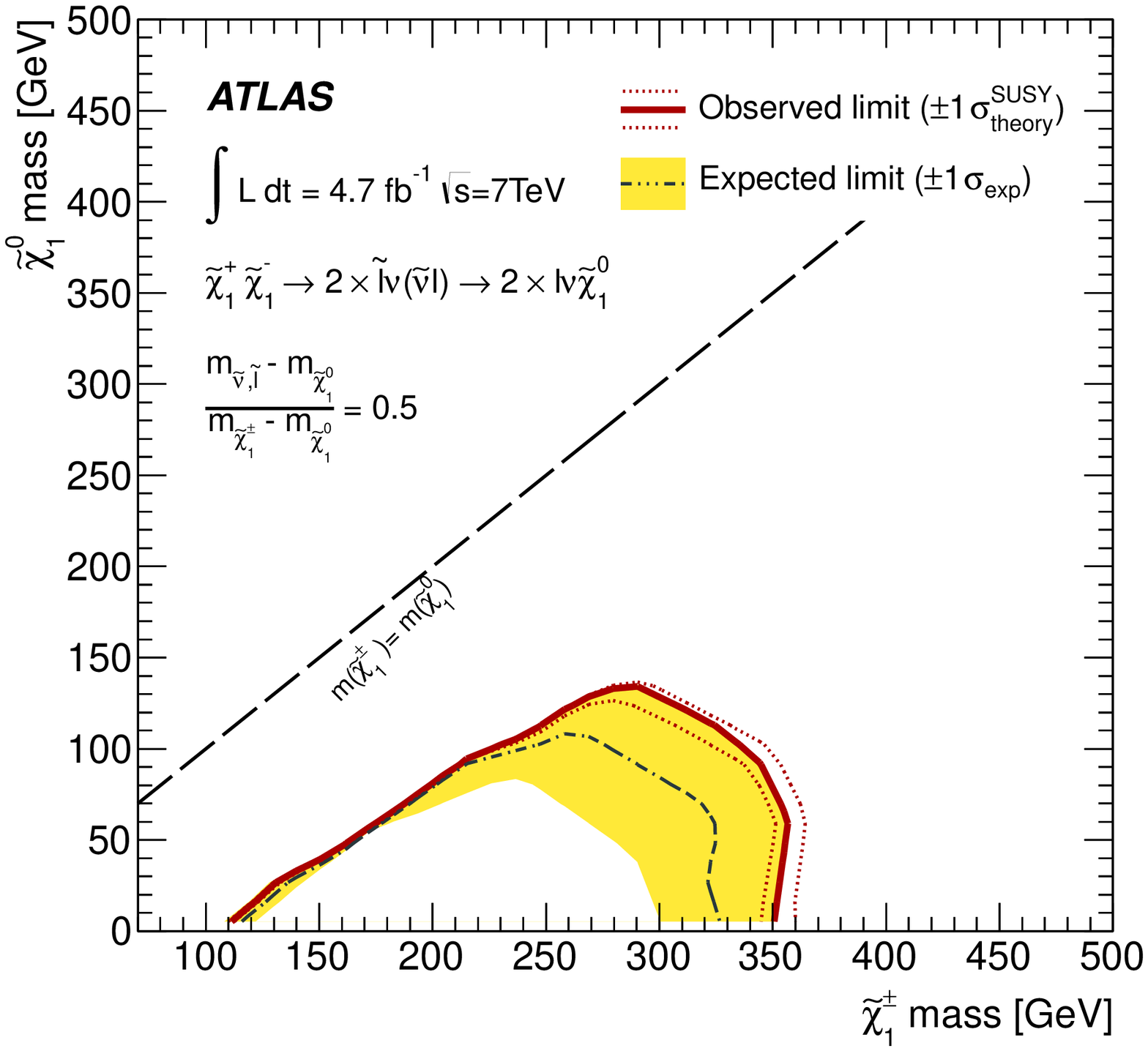}
\caption[*]{Summary of CMS gaugino mass limits in a number of
simplified models assuming $\chinopm \ninotwo$ pair
production~\cite{cmsgauginos} 
(left), and ATLAS gaugino mass limits assuming a simplified
model of chargino pair production and
decay~\cite{atlas2lgauginos} (right).
}
\label{fig:gauginoresults}
\end{center}
\end{figure}

The CMS results for searches for $\chinopm \ninotwo$ pair
production~\cite{cmsgauginos} are summarized in 
Fig.~\ref{fig:gauginoresults} (left). The simplified model used
assumes $m_{\chinopm} = m_{\ninotwo}$ and, for the light
slepton scenario, $m_{\slepton} = 0.5 m_{\ninoone} +
0.5 m_{\chinopm}$. For the heavy slepton scenario with 
$W/Z$-like decays, chargino limits extend up to
$230 \GeV$ for massless LSPs, whereas for enhanced leptonic 
branching fractions with light intermediate sleptons mass 
limits are higher, up to chargino masses of $500 \GeV$ 
for massless LSPs in the light $\sleptonL$ scenario.
The ATLAS analyses~\cite{atlas3lgauginos,atlas2lgauginos} lead
to very similar conclusions. The results of ATLAS limits on
chargino-pair production in the light slepton scenario are
shown in Fig.~\ref{fig:gauginoresults} (right). A by-product of the
ATLAS gaugino analyses are limits on slepton pair 
production~\cite{atlas2lgauginos}.

Both ATLAS and CMS have now proven sensitivity for realistic
gaugino production and decay scenarios, and have set the first
LHC limits on gauginos within simplified models. This is an
important milestone, a relatively recent one, and one that
promises a lot for the near future.

\section{SUSY Escape Routes}

In view of the null results obtained in all searches,
it is worthwhile to consider whether SUSY could have escaped
all these analyses, and how. There are certainly a number of
ways that can be imagined, and dedicated analyses exist or are
under development to close these holes.

\subsection{Compressed spectra}

As shown in Fig.~\ref{fig:simplified}, limits on squark and
gluino masses in the $\TeV$ range are only valid for light
neutralinos. For neutralinos with masses above some $400 \GeV$,
squark and gluino mass limits collapse. This is due to the
fact that the analyses look for energetic jets and significant
missing transverse momentum, but if the missing particle is
massive itself, jets carry less energy, and the missing momentum
is significantly reduced. This poses a problem to the analyses:
lowering cuts will lead to much enlarged backgrounds, in fact
these backgrounds lead to problems triggering on such events.
One development in this area is the use of ``delayed triggers'',
or ``parked data'': these are data streams collected with
triggers with lower thresholds that will only be processed
after the completion of the 2012 data taking. Results of analyses
using these triggers can be expected in 2013.

\subsection{Long lived particles}

SUSY particles need not decay promptly; for various reasons they
could have a sizable lifetime and decay anywhere in the detector,
or even be semi-stable and leave the detector before they decay.
Dedicated analyses for long-lived particles have been developed
by ATLAS and CMS.

R-hadrons are objects containing a long-lived gluino or squark,
that hadronizes to an R-meson or R-baryon. Such R-hadrons may
be produced in $pp$ collisions and be absorbed by detector material
such as the calorimeters of ATLAS and CMS, and then decay much
later, uncorrelated with any beam activity. CMS have searched for
such decays during gaps between LHC beam crossings~\cite{cmsstoppedgl}.
Using a data set in which CMS recorded an integrated luminosity of
 4.0 fb$^{-1}$, and a search interval corresponding to 246 hours of trigger 
live time, $12$ events are observed, with a mean background prediction 
of $8.6 \pm 2.4$ events. Limits are derived at 95\% confidence level 
on long-lived gluino and stop production, over 13 orders of magnitude 
of R-hadron lifetime.

Particles such as R-hadrons, or semi-stable sleptons, that traverse
the detector before they decay, appear like stable massive charged
particles. They have high transverse momentum, but could be slow,
and can be searched for by measuring their specific ionization or
their time-of-flight through the detector~\cite{exotictalk}.
Non-observation of a signal above SM background leads to limits
on ``stable'' gluinos of approximately $1 \TeV$, on stable stops
of $700 \GeV$, and on a stable stau of $300 \GeV$.

\begin{figure}[!thb]
\begin{center}
\includegraphics[width=6cm]{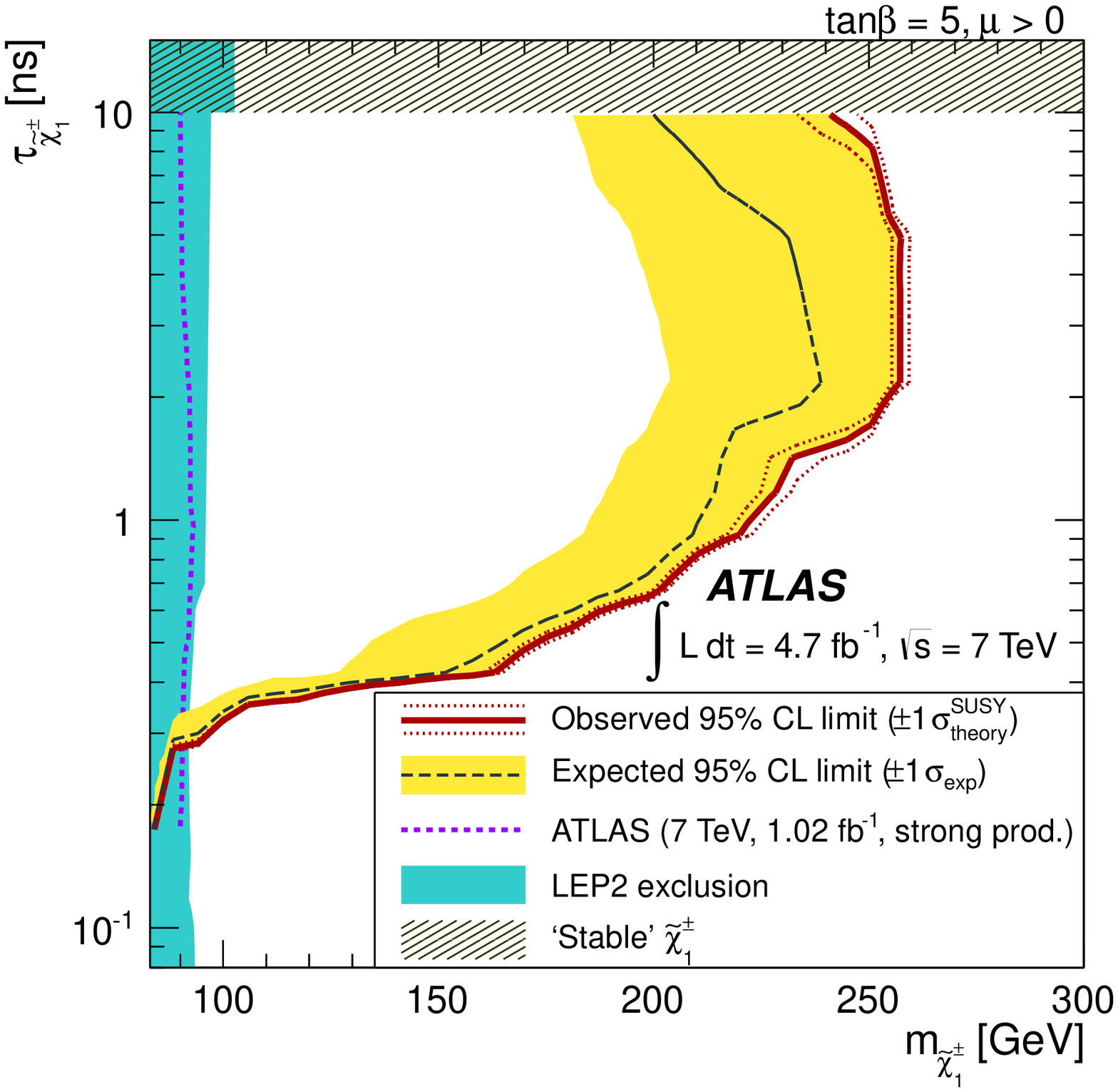}
\includegraphics[width=6cm,height=6cm]{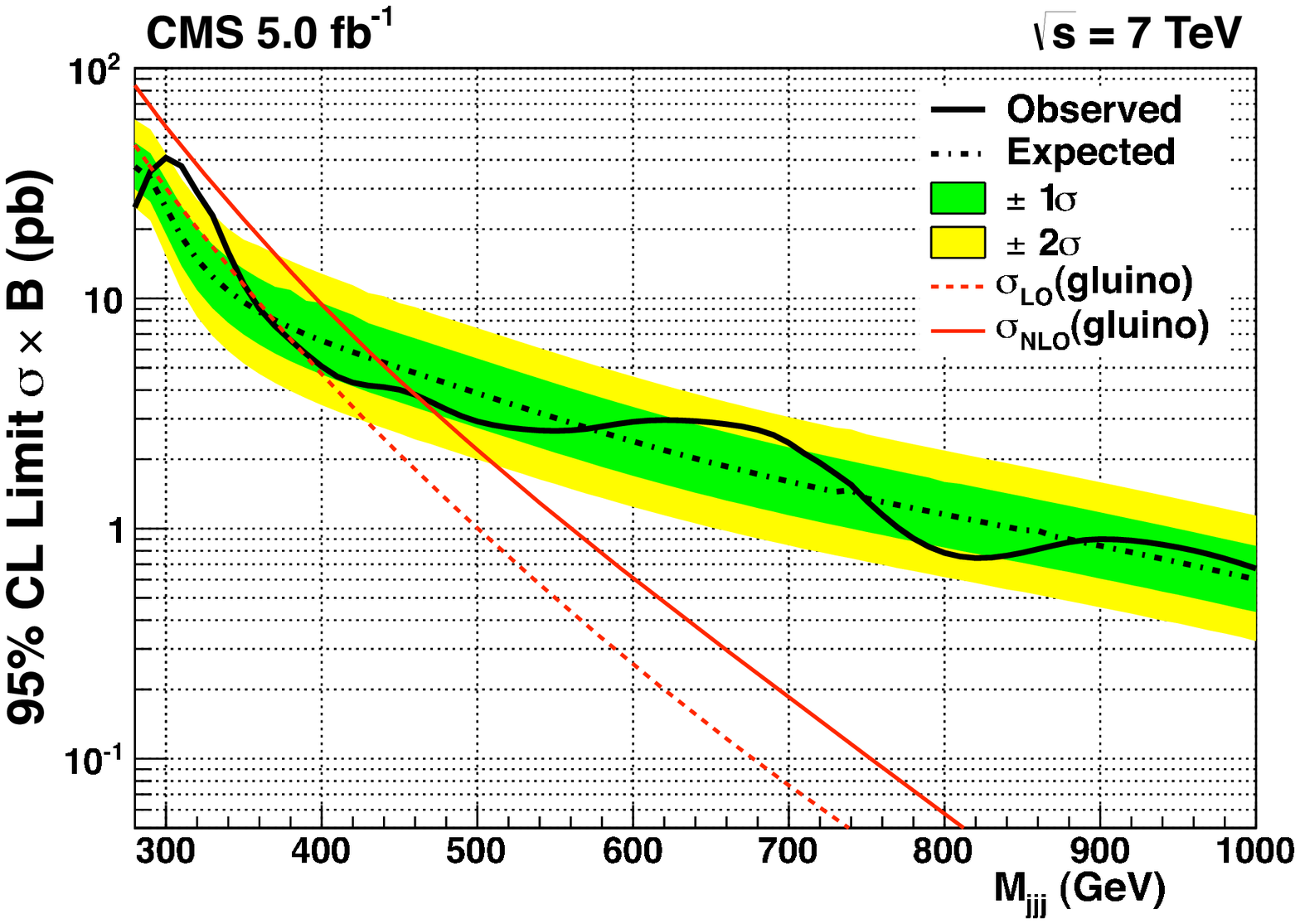}
\caption[*]{ATLAS chargino lifetime limits in the AMSB model
from a disappearing track analysis~\cite{disapptrack} (left), 
and CMS limits on the mass of RPV decaying 
gluinos~\cite{cmsthreejets} (right).
}
\label{fig:escape}
\end{center}
\end{figure}

ATLAS has searched for ``disappearing'' tracks~\cite{disapptrack},
in a scenario motivated by anomaly-mediated SUSY breaking (AMSB) models 
in which
charginos are almost mass-degenerate with the lightest neutralino.
Chargino decay is suppressed: only the decay channel to neutralino plus
soft pion is open. The soft pion track, which originates in a vertex
far away from the interaction point, is easily missed in the event
reconstruction, leading to the signature of a high $\pt$ track that
seems to disappear. ATLAS selects events with tracks with $\pt > 10$
$\GeV$ with good track quality in the inner tracker, but less than
five hits in the outer segment of the transition radiation tracker (TRT).
Backgrounds arise from tracks interacting with the TRT material and
from mismeasured tracks; no excess is seen above the background
predictions. Resulting exclusion limits in the chargino mass-lifetime
plane are shown in Fig.~\ref{fig:escape} (left).

\subsection{R-parity violation}

If R-parity is violated, the following additional terms enter in the
superpotential:
\begin{displaymath}
W_{\mathrm{RPV}} = \lambda_{ijk} \hat{L}_i \hat{L}_j \hat{E}_k^C +
\lambda'_{ijk} \hat{L}_i \hat{Q}_j \hat{D}_k^C +
\lambda''_{ijk} \hat{U}_i^C \hat{D}_j^C \hat{D}_k^C +
\varepsilon_i \hat{L}_i \hat{H}_u
\end{displaymath}
where $\hat{L}$ and $\hat{Q}$ are the leptonic and quark superfields
containing the left-handed fermions and their scalar SUSY partners,
$\hat{E}$, $\hat{U}$ and $\hat{D}$ are the charged leptonic, up-type
quark and down-type quark superfields containing the right-handed
fermions and their SUSY partners, and $\lambda,\lambda',\lambda''$
are R-parity violating couplings. The last term in the superpotential
 is the bilinear term
involving the Higgs field, and will not be discussed here.
Non-zero values of $\lambda$ lead to multilepton final states, non-zero
values of $\lambda''$ lead to multijet final states (or resonances),
and non-zero $\lambda'$ leads to mixed leptonic-hadronic final states
not unlike those of leptoquark decays.

CMS has searched for multilepton final states~\cite{cmsmultilepton},
with three or four leptons, including either no, one or two $\tau$
lepton(s). CMS has binned their data in bins of $\HT$ and $\met$,
and generally observe good agreement between data and SM predictions.
CMS also looks for three-jet resonances that could arise from
R-parity-violating gluino decays~\cite{cmsthreejets}. In events
with significant $\HT$ and large jet multiplicity, the distribution
of three-jet invariant mass is plotted and fit with an Ansatz for
a background function; the residuals of the fit show no outliers and limits
on gluino masses assuming RPV gluino decays
are set, as shown in Fig.~\ref{fig:escape} (right).

Recent ATLAS results on R-parity violating SUSY searches include a search
for non-resonant $e + \mu$ production from quark-quark scattering,
mediated by $t$-channel top squark exchange~\cite{atlasemu}, 
and a search for displaced, massive vertices with high track multiplicity
accompanied by a high $\pt$ muon, as a sign of neutralino 
decay~\cite{atlasdispvertex}. Both analyses observe no excess of
events above the SM prediction, and put limits on $\lambda'$.

\subsection{Beyond the MSSM}

Models with a particle content beyond that of the {\it minimal}
supersymmetric extension of the SM are of interest as well.
In such models it is often easier to generate a mass for the
lightest Higgs boson as large as $125 \GeV$~\cite{higgs} than in the MSSM.
Beyond the MSSM models of SUSY could have final states for which
current SUSY searches are not (yet) optimized.

CMS has searched for a final state of two photons, four or more
hadronic jets and low $\met$, motivated by ``stealth SUSY''
models~\cite{cmsstealth}.
ATLAS has searched for pair production of scalar colour-octets,
(i.e. scalar gluons), decaying to two pairs of 
jets~\cite{scalargluon}. Both analyses observe no excess of events
above the SM expectation.

\section{Conclusion and Outlook}

\begin{figure}[!thb]
\begin{center}
\includegraphics[width=8cm]{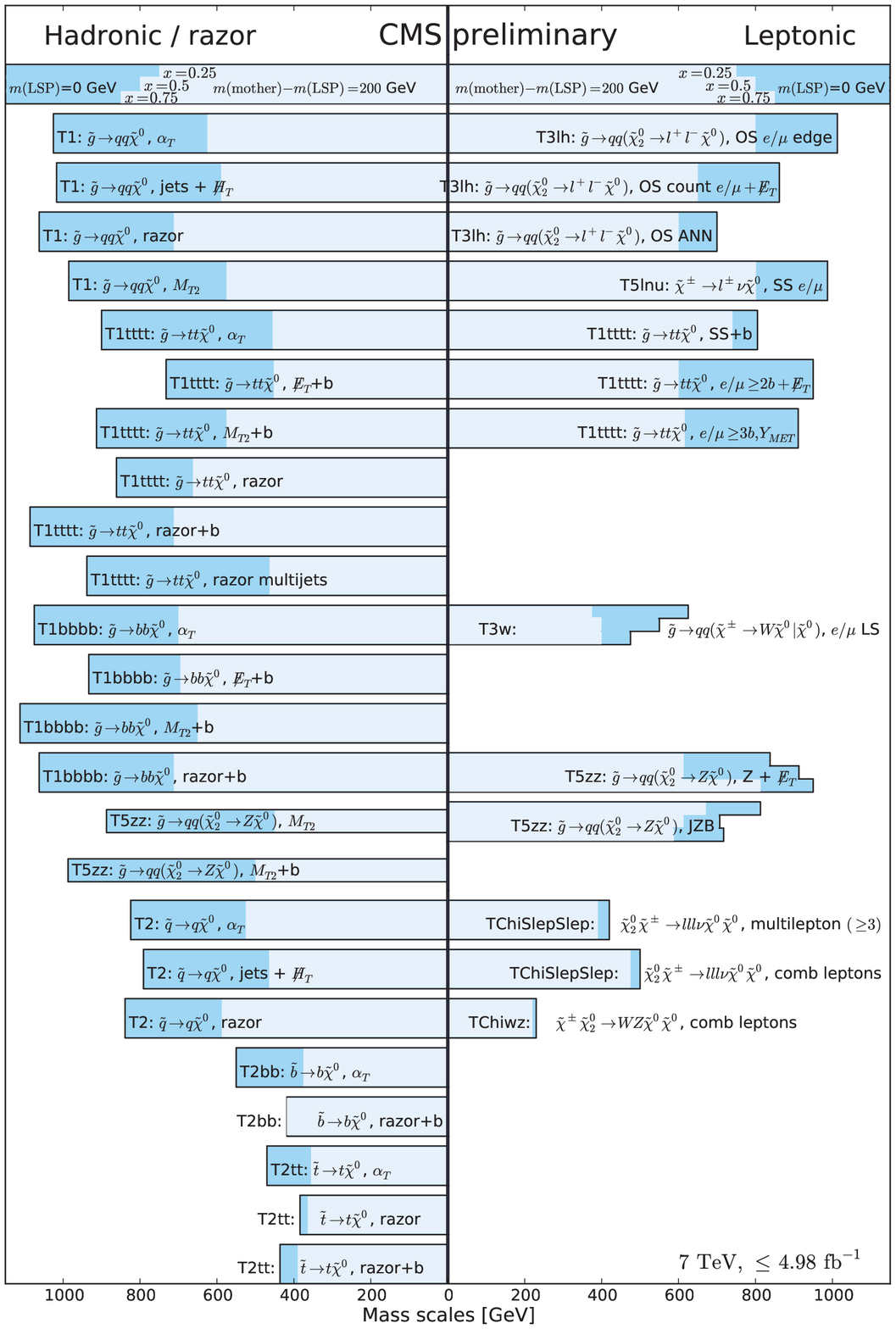}
\includegraphics[width=10cm]{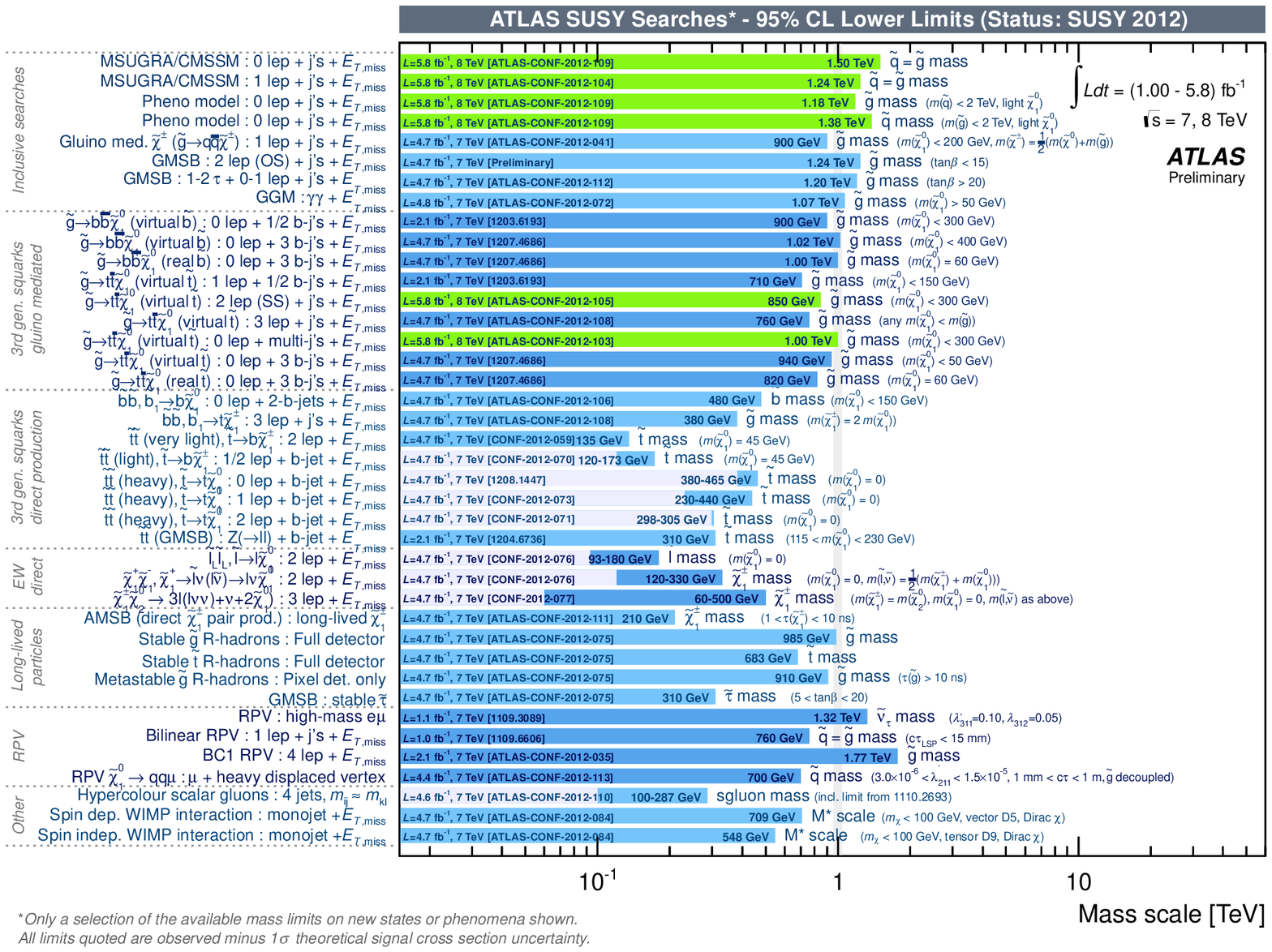}
\caption[*]{Summary of CMS SUSY search results interpreted in simplified
models (top)~\cite{cmssmssummary} and of various ATLAS SUSY search 
results (bottom)~\cite{atlascombinedsummaryplots}.
}
\label{fig:summary}
\end{center}
\end{figure}

For ATLAS and CMS, searches for SUSY are high priority, and this is
reflected in the very large number of public notes and papers.
Some of these results are summarized in 
Fig.~\ref{fig:summary}~\cite{cmssmssummary,atlascombinedsummaryplots}.

Unfortunately, none of the analyses have observed a significant
deviation from SM predictions, leading to constraints on SUSY
models. For strong production, limits on squarks and gluinos
are in the $\TeV$ region for MSUGRA/CMSSM, and similar limits hold
in simplified models unless spectra are compressed.
Recently, emphasis has shifted somewhat towards dedicated searches
for third generation squarks and for gauginos. First results for
such searches have appeared, proving that sensitivity exists:
the attack on ``natural SUSY'' has now truely started.
More results can be expected in the near future, making use of
the full 2012 LHC data set.
Further emphasis can also be expected on analyses trying to 
close the gaps through which SUSY can escape: final states
with low $\met$, or with long-lived particles. After the
2013-2014 LHC shutdown, it is expected that the LHC beam energy
will be increased to at least $6.5 \TeV$, and with the expected
large integrated luminosity that will be collected, a whole
new era of SUSY searches will start in 2015.

\end{document}